\documentclass[fleqn,usenatbib]{mnras}

\usepackage{newtxtext,newtxmath}

\usepackage[T1]{fontenc}
\usepackage{ae,aecompl}

\usepackage{graphicx}	
\usepackage{amsmath}	

\title[He~{\sc i}~1.08~$\mu$m absorption in NGC~5548]{Broad He~{\sc i}~1.08~$\mu$m absorption from the obscurer in the active galaxy NGC~5548}

\author[C. Wildy et al.]{
Conor Wildy,$^{1}$\thanks{E-mail: wildy@cft.edu.pl}
Hermine Landt,$^{2}$\thanks{E-mail: hermine.landt@durham.ac.uk}\thanks{Visiting Astronomer at the Infrared Telescope Facility, which is operated by the University of Hawaii under contract NNH14CK55B with the National Aeronautics and Space Administration.}
Martin J. Ward,$^{2}$\footnotemark[3]
Bozena Czerny,$^{1}$
Daniel Kynoch$^{3}$\footnotemark[3]
\\
$^{1}$Center for Theoretical Physics, Polish Academy of Sciences, Al. Lotnik\'ow 32/46, 02-668 Warsaw, Poland \\
$^{2}$Centre for Extragalactic Astronomy, Department of Physics, Durham University, South Road, Durham DH1 3LE, UK \\
$^{3}$Astronomical Institute, Academy of Sciences, Bo\v{c}n\'{i} II 1401, 14131 Prague, Czech Republic
}

\date{Accepted . Received ; in original form }

\pubyear{2020}

\begin{document}
\label{firstpage}
\pagerange{\pageref{firstpage}--\pageref{lastpage}}
\maketitle

\begin{abstract}
The nucleus of the active galaxy NGC~5548 was the target of two intensive spectroscopic monitoring campaigns at X-ray, ultraviolet (UV), and optical frequencies in 2013/14. These campaigns detected the presence of a massive obscuration event. In 2016/17, Landt et al. conducted a near-IR spectroscopic monitoring campaign on NGC~5548 and discovered He~{\sc i}~1.08~$\mu$m~absorption. Here we decompose this absorption into its components and study its time variability. We attribute the narrow He~{\sc i} absorption lines to the warm absorber and, as for the newly appeared low-ionization warm absorber lines in the UV, their presence is most likely due to a reduction in ionization parameter caused by the obscurer. The observed variability of the narrow He~{\sc i} absorption is consistent with what is expected for the warm absorber. Most importantly, we also detect fast, broad  He~{\sc i} absorption, which we attribute to the obscurer. This He~{\sc i} broad absorption, which is indicative of a high-column density gas, is unsaturated and variable on time-scales of a few months. The observed variability of the obscurer is mainly due to changes in ionization, although density changes also play a role. We test the physical cycle model of Dehghanian et al. which proposes that helium recombination can account for how the obscurer influences the physics of the warm absorber gas. Our results support their model, but also indicate that the reality might be more complex.
\end{abstract}




\section{Introduction}

Active Galactic Nuclei (AGN) are luminous and persistent energy sources at the centres of galaxies, powered by accreting supermassive black holes. Outflows from AGNs limit the rate of black hole growth and can significantly influence star formation in their host galaxy by depositing their kinetic luminosity into the surrounding environment via the so-called feedback process \citep{springel05,martinnavarro18}. The outflowing material, if located along the line-of-sight to the AGN, can appear as blue-shifted absorption lines in AGN spectra, offset from the systemic emission line velocity by sometimes more than 10~percent of the speed of light \citep{pounds03,tombesi10}. Such absorption lines have been observed across a wide range of the electromagnetic spectrum, including near-infrared (near-IR), optical, ultraviolet (UV) and X-ray wavelengths. The outflowing ions producing absorption have been observed in a variety of ionization states, implying that a corresponding diversity of ionizing environments exists both within and between objects \citep{netzer96}. 

Variability of AGN absorption line strength is a commonly observed phenomenon and has been measured over timescales ranging from days to years \citep{lundgren07,capellupo11,grier15}. Possible explanations for variability include changes to the covering fraction of the emission region, due to motion of the absorber, and changes in the ionization state of the gas, due to variations in the incident ionizing radiation \citep{gibson08,wang15}. Distinguishing between these cases is possible through the use of photoionization simulations and measuring the level of coordination in variability between different ions and across large velocity separations.

Atomic helium in the 2$^{3}$S state is considered metastable, as its theoretical average lifetime is 2.2~hr \citep{drake71}. This state can be useful for studying high column density outflows since it has a relatively low abundance, making it less prone to absorption-line saturation compared to more-commonly used ions such as C~{\sc iv} and Si~{\sc iv}. Within AGN outflows, metastable helium is almost entirely formed from recombinations of He~{\sc ii} due to it being 19.8~eV above the true ground state of He~{\sc i}, which is much greater than the absorber electron temperature. It therefore only forms in ionizing environments suitable for the generation of He~{\sc ii}. The transitions of metastable helium form a triplet at vacuum wavelengths of 3189~\AA{} (UV), 3890~\AA{} (optical) and 10\,833~\AA{} (near-IR). The near-IR component is relatively easy to measure due to its strength and relative lack of blending with neighbouring emission lines. The effectiveness of using this metastable state as a means of understanding AGN outflows has been demonstrated previously \citep{leighly11,wildy16}. 

The active core of the galaxy NGC~5548 is one of the best-documented AGN due to both its relative proximity to Earth, providing high line fluxes, and its strong continuum variability. This has made it a prime target for reverberation-mapping studies \citep[e.g.][]{peterson02,bentz07} allowing determination of the black hole mass through broad emission line time delays. X-ray observations prior to 2013 show substantial emission in the soft band \citep[e.g.][]{done95,mckernan07}, which is characteristic of an unobscured source and in agreement with its optical spectral properties such as broad emission lines and strong accretion disc continuum emission. It is also known to harbour persistent outflows due to highly ionized species such as N~{\sc v} and C~{\sc iv}, which manifest as a series of narrow UV absorption lines of full-width at half-maximum (FWHM) up to a few hundred km~s$^{-1}$. These lines have been seen in observations dating from the 1990s onwards \citep{mathur99,crenshaw99,crenshaw03,arav15} and are sometimes ascribed to an X-ray warm absorber (WA). The UV and X-ray absorption lines from the WA are most likely produced in gas of lower and higher ionization, respectively, implying a stratification in the outflow. The WA is known to exist at the parsec scale and its properties have been observed to vary \citep{steenbrugge05,ebrero16}.

Two large observing campaigns during the summer of 2013 and the first half of 2014 monitored light-curves and the spectral behaviour of NGC~5548 using high-energy telescopes including \emph{XMM-Newton}, \emph{NuSTAR}, \emph{Chandra} and \emph{Swift}, along with optical/UV data obtained from the \emph{Hubble Space Telescope} and ground-based observatories \citep{kaastra14,digesu15,mehdipour15,mehdipour16,derosa15,mathur17}. These campaigns revealed considerable soft X-ray obscuration not present in previous observations, together with previously unseen UV narrow absorption lines resulting from low- to moderate-ionization species such as Si~{\sc ii}, Si~{\sc iv} and C~{\sc ii} \citep{kaastra14,kriss19}. New broad absorption troughs spanning a few thousand km~s$^{-1}$ from a fast outflow were also seen within the UV spectra and attributed to the X-ray obscurer. The appearance of lower-ionization narrow lines in the WA has been explained by a reduction in the ionization parameter of the outflowing material, most likely caused by absorption of ionizing photons from the nucleus by the obscurer, which is located much further in \citep{dehghanian19a,kriss19}. Variability monitoring of the X-ray obscurer by \citet{cappi16} reported the most recent measurements of its physical properties and described it as consisting of one moderately ionized component and one cold neutral component. Variability of the UV narrow absorption lines was reported previously by \citet{crenshaw03,crenshaw09}, but only recently \citet{kriss19} studied also the UV variability of the obscurer and reported a decrease in strength over a period of $\sim 6$~months. They suggested that the obscurer is a powerful wind off the accretion disc, a model which has recently been elaborated on by \citet{dehghanian19a, dehghanian19b}. The main arguments for the interpretation of the obscurer as an accretion disc wind were: (i) the observation of a ``line holiday'', i.e. a de-correlation of the variability of the line flux from that of the ionizing extreme UV (EUV) continuum flux for part of the campaign, not only for the narrow absorption lines but also for the broad emission lines \citep{goad16}; (ii) an anti-correlation between the strength of the soft X-rays and the equivalent width of the broad UV absorption troughs; and (iii) no strong changes in the velocity widths of the broad UV absorption troughs. The first observation indicates that the location of the obscurer is between the continuum source and broad-line region (BLR) and that the obscurer has an axisymmetric geometry rather than being a single cloud (as could be the case if the ``line holiday'' was observed only for the narrow absorption lines). The latter interpretation is supported also by the third observation. The second observation points to an outflowing obscurer.  

In their one-year long near-IR (NIR) reverberation campaign on NGC~5548 during 2016/17, \citet{landt19} discovered He~{\sc i} absorption at 10\,833~\AA. Here we analyse the different components of the He~{\sc i} absorption and study their time variability. The appearance of narrow He~{\sc i} absorption lines is most likely related to the appearance of low-ionization WA lines in the UV. However, most interestingly, we detect also fast, broad  He~{\sc i} absorption, which we attribute to the obscurer. The paper layout is as follows: Section~2 details the observations; Section~3 describes the analysis leading to the identification of absorption components and measurements of their variability; Section~4 provides discussion of the findings from Section~3 in more detail, and Section~5 provides a summary of the main results of this work. All laboratory line wavelengths quoted in this paper are vacuum wavelengths, and velocities are defined as negative if in the blue-shifted (outflowing) direction.

\section{The near-IR spectroscopy}

\citet{landt19} observed NGC~5548 between 2016 August and 2017 July with the recently refurbished SpeX spectrograph \citep{rayner03} at the NASA Infrared Telescope Facility (IRTF), a 3~m telescope on Maunakea, Hawaii. The aim of this near-IR spectroscopic reverberation campaign was to measure the lag time of the hot dust in the obscuring torus together with the dust temperature, as well as the variability of emission lines such as those from the coronal line region and BLR. However, a serendipitous discovery was the presence of absorption in the He~{\sc i}~$\lambda 10830$ broad emission line (see their Fig. 1). 

In summary, the campaign achieved a total of 18 near-IR spectra with an average cadence of about ten days, excluding the 3.5-month period (Sep to mid-Dec) when the source is unobservable. They used the SXD cross-dispersed mode ($0.7-2.55~\mu$m) and a $0.3'' \times 15''$ slit oriented at the parallactic angle, resulting in an average spectral resolution of $R=2000$ or FWHM of $\sim 150$~km~s$^{-1}$. As discussed in \citet{landt19}, the narrow slit and the optimal spectral extraction rendered the flux contamination from the host galaxy negligible. The spectra have a high signal-to-noise ($S/N$) ratio with an average continuum $S/N \sim 60$ in the $J$~band, where the He~{\sc i} line is observed.

We selected eight observations from the data of \citet{landt19} (Table \ref{tab:obs}). The selection was made on the basis of both a high continuum $S/N$ and an even spread over the duration of the campaign. In particular, we excluded spectra that were noisy in the wavelength region close to the atmospheric absorption window between the $J$ and $H$~bands, which partially overlaps with the broad Pa$\beta$ line on its longer wavelength (red) side. All selected spectra show obvious absorption blue-ward of the He~{\sc i} line and are hereafter referred to as the ``absorbed spectra''. A single observation from 2006 June published by \citet{landt08}, which appears to have unabsorbed He~{\sc i}, was used to provide emission templates for the more recent absorbed spectra. The 2006 spectrum was obtained through a wider slit ($0.8''$) resulting in a lower spectral resolution of $R\sim800$. By degrading the spectral resolution of the new spectra to this value, we checked that the current He~{\sc i} absorption would have been detectable if it had been present then. 

Initially, small additive corrections (up to 1.6~\AA{}) were made to each spectrum to align skylines with their laboratory wavelengths. Subsequently the spectra were transformed to the object rest-frame for all further analysis. Differences in flux calibration among absorbed spectra were corrected using the photometric scaling factors listed in \citet{landt19} (see their Table 1). In summary, these were obtained by first using near-IR integral field unit (IFU) observations to correct for slit losses and then matching the profiles of the strong narrow forbidden emission line [S~{\sc iii}]~$\lambda$9533 in the different spectra. Due to the physical size of the emitting region, the [S~{\sc iii}]~$\lambda$9533 line is not expected to vary on the timescales of several months or even a few years. The continuum signal-to-noise (S/N) reported in Table \ref{tab:obs} is the average value calculated in two continuum-fitting windows either side of the He~{\sc i} blend, having wavelength ranges of 9700--9800~\AA{} and 12\,000--12\,100~\AA{}.

\begin{table*}
\begin{center}
\caption{List of observations selected from \citet{landt19}}
\begin{tabular}{l l l l l l c l c}
\hline\hline Observational & Observation & HJD & Exposure & Airmass & Seeing & Slit width & Continuum & Correction \\
 Epoch & Date & -2,400,000 & (s) & & (arcsec) & (arcsec) & S/N$^{\dagger}$ & factor \\
\hline
 *&2006 Jun 12&&16~$\times{}$~120&1.103&0.9&0.8&49&\\
 1&2016 Aug 2 &57602.76&16~$\times{}$~120&1.180&0.7&0.3&62&0.85\\
 2&2016 Aug 11&57611.74&16~$\times{}$~120&1.209&0.6&0.3&51&1.03\\
 3&2016 Dec 21&57744.08&16~$\times{}$~120&2.184&0.4&0.3&38&0.78\\
 4&2017 Feb 15&57800.11&12~$\times{}$~120&1.006&0.8&0.3&37&0.73\\
 5&2017 Feb 24&57809.10&16~$\times{}$~120&1.005&0.6&0.3&53&1.15\\
 6&2017 Mar 17&57830.01&16~$\times{}$~120&1.027&0.7&0.3&38&1.00\\
 7&2017 Mar 22&57834.99&18~$\times{}$~120&1.030&0.5&0.3&57&0.77\\
 8&2017 May 9 &57882.85&16~$\times{}$~120&1.041&0.6&0.3&44&0.89\\
\hline
\hline
\end{tabular}
\label{tab:obs}
\end{center}
\parbox[]{16cm}{*Used as a reference spectrum for the unabsorbed He~{\sc i} broad emission line profile.\\
$^{\dagger}$Measured as detailed in the main text.\\}
\end{table*}

\section{Spectral analysis}

\subsection{The unabsorbed spectrum}

A continuum constructed as the sum of a power-law (corresponding to the accretion disc emission) and a blackbody (corresponding to the dust emission), without any galaxy contribution, gave a good fit to line-free continuum regions in the absorbed spectra. However, for the unabsorbed reference spectrum from 2006 June, which was taken with a wider slit (0.8"), the addition of galaxy starlight and hence a galaxy template was required. For this purpose we used a Sa-type spiral template from the SWIRE library \citep{polletta07}. We performed both the continuum fits and the line fits described in detail below with the \emph{Specfit} package within the \emph{Image Reduction and Analysis Facility} (IRAF) software \citep{iraf93}.
 
\begin{figure}
	\includegraphics[width=\columnwidth]{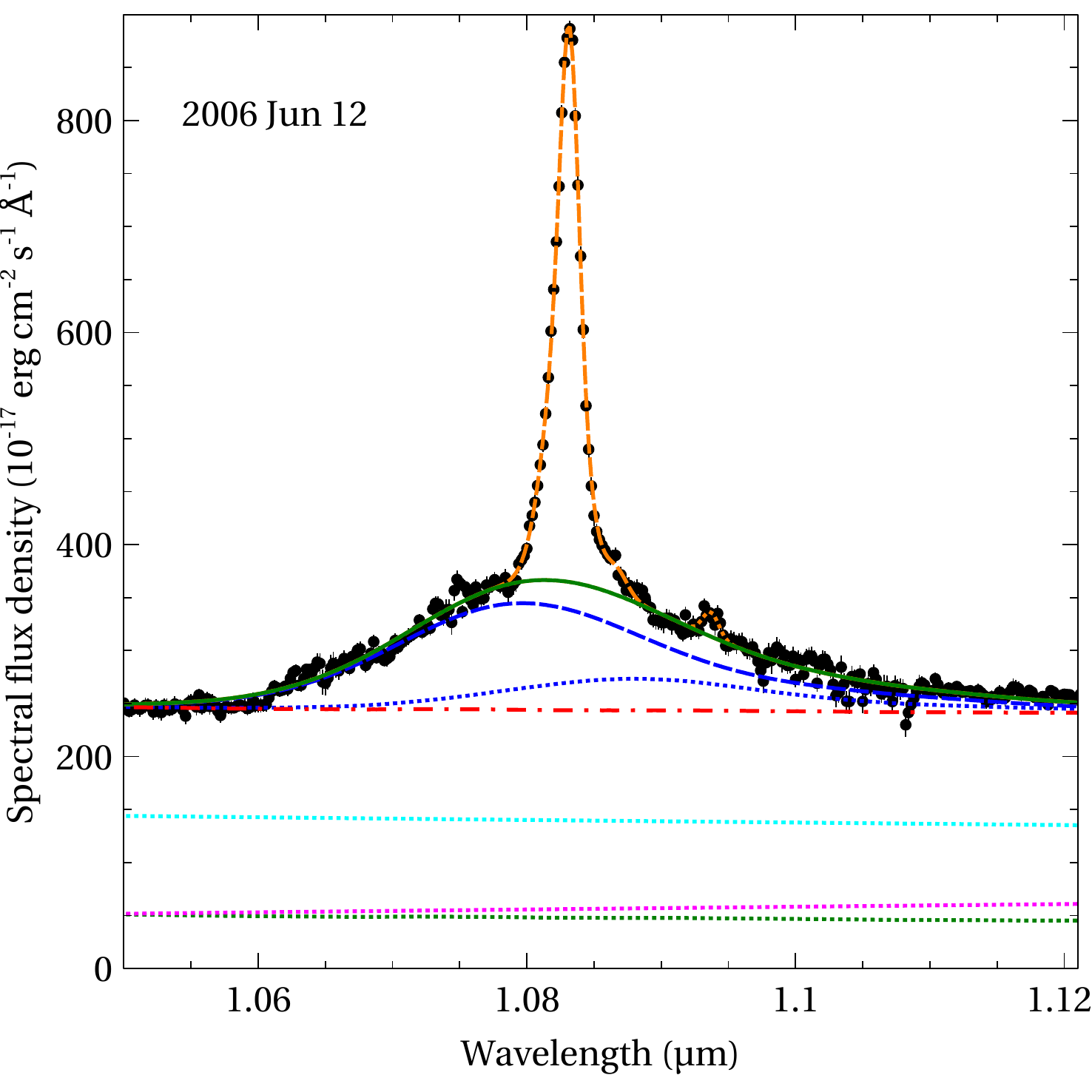}
    \caption{Spectral region of the He~{\sc i} blend from the 2006 June (reference) spectrum (black circles). The red dotted-dashed line represents the total continuum formed from the sum of disc (dotted cyan line), dust (magenta dotted-line) and galaxy starlight (green dotted-line) components. Blended line emission excluding narrow components is represented by the solid green line. Blue lines represent the broad components of He~{\sc i} (dashed) and Pa$\gamma$ (dotted). Orange lines represent the narrow components of He~{\sc i} (dashed) and Pa$\gamma$ (dotted).}
    \label{fig:jun06}
\end{figure}

\begin{figure}
	\includegraphics[width=\columnwidth]{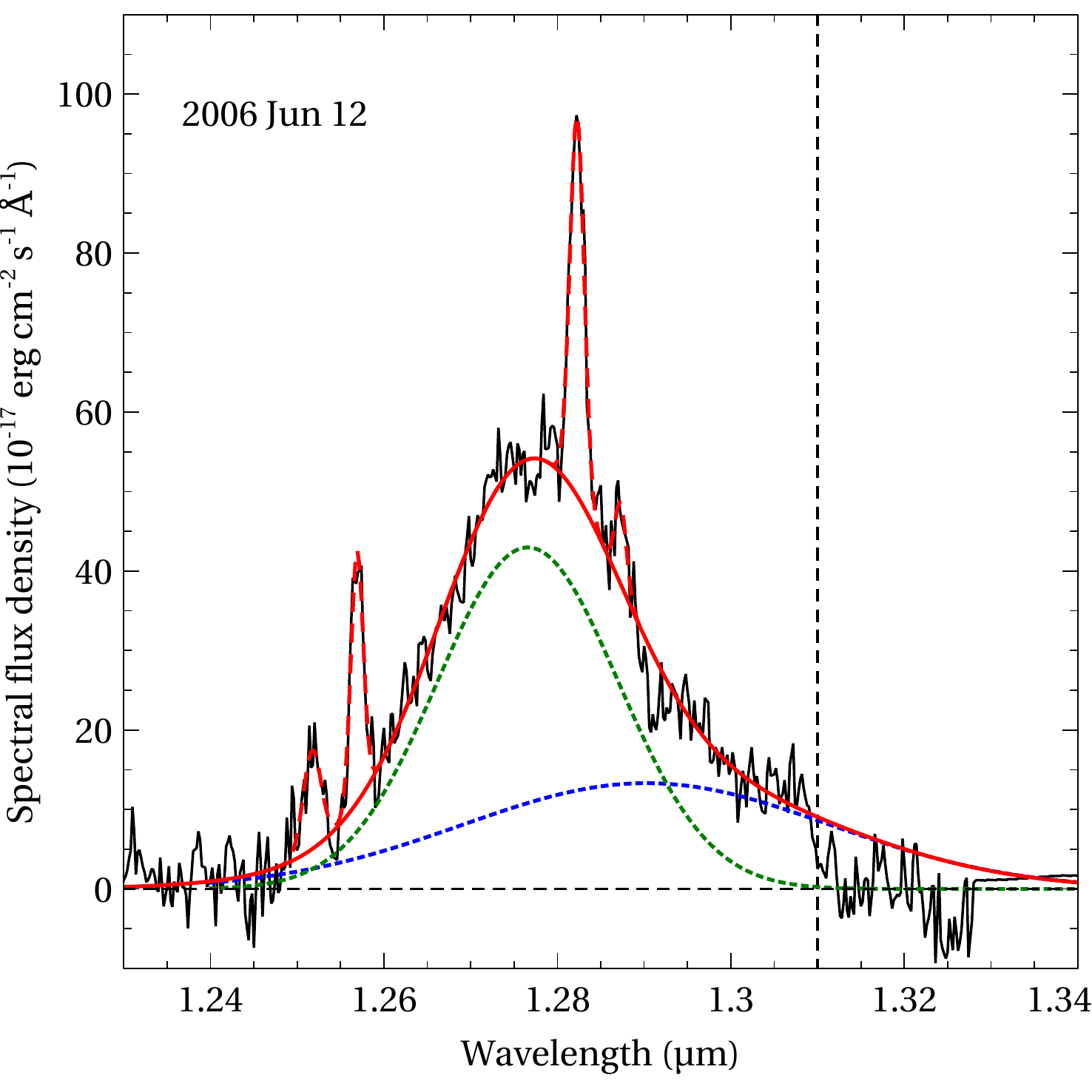}
    \caption{Continuum-subtracted Pa$\beta$ blend from the reference spectrum (black line). The green and blue dotted lines are Gaussian models which sum to give the broad Pa$\beta$ profile (solid red line). Dashed red lines show the blended narrow lines. The fit was restricted to $\lambda$$\leq$13100 due to poor fitting at longer wavelengths caused by proximity to the upper boundary of the J-band transmission window.}
    \label{fig:pab}
\end{figure}

In order to obtain templates for the broad and narrow components of the unabsorbed He~{\sc i} $\lambda$10\,833 emission blend, we performed a multi-component simultaneous fit to the blend observed in the reference spectrum, the result of which is shown in Fig.~\ref{fig:jun06}. Due to the high velocity width and hence wavelength spread of the broad lines in NGC~5548, the He~{\sc i} line has substantial blending with Pa$\gamma$ at 10\,941~\AA{}. Therefore, we generated a template for broad Pa$\gamma$ using broad Pa$\beta$, which is not blended with any other broad lines \citep{landt08}. Broad Pa$\beta$ was isolated by generating Gaussian models to approximate the blended narrow lines of narrow Pa$\beta$, [S~{\sc ix}]~$\lambda$12\,523, [Fe~{\sc ii}]~$\lambda$12\,570 and an unknown line at 12\,873~\AA{}, and then subtracting them off. The broad Pa$\gamma$ template was then created by fitting the isolated broad Pa$\beta$ line with two Gaussian profiles followed by transformation of this fit to the wavelength of Pa$\gamma$. No physical significance is attributed to the individual Gaussian components as they are only used to approximate (in sum) the total broad Pa$\beta$ profile. All modelled components contributing to the Pa$\beta$ blend are illustrated in Fig.~\ref{fig:pab}.

The lower limit to the strength of the broad Pa$\gamma$ template relative to Pa$\beta$ was restricted to that produced by Case B recombination in a low density gas at 10\,000~K, as listed in \citet{osterbrock06}, after a small correction for internal reddening was applied \citep{gaskell07}. High density gases could however produce a less steep Paschen decrement, so the upper limit was set equal to the Pa$\beta$ strength. Additionally, a single Gaussian was generated to represent the relatively weak narrow component of Pa$\gamma$. A double-Gaussian profile was used to model the broad component of He~{\sc i}, while the narrow component was constructed using a triple-Gaussian profile. As in the case of Pa$\beta$, no physical significance is attributed to individual Gaussian components in these multi-Gaussian profiles. No significant deviation from the total emission model was found at wavelengths where the possibility of broad Fe~{\sc ii} emission exists, so it was not added to the model profile.
 
Subsequently, a mean spectrum was constructed from the (continuum-subtracted) absorbed spectra. After being adjusted to match the higher effective resolution of the absorbed spectra, an attempt was made to use the He~{\sc i} broad and narrow components from the reference spectrum as (fixed-shape) templates for fitting the same components in the mean spectrum. This followed a similar procedure to that used for the 2006 June reference spectrum, differing by the use of a two-step process. The first step involved fitting to the red-side of the blend to avoid the blue-shifted absorption. Second, any regions from the blue-side of the blend where the emission model appeared significantly below the data points were additionally designated as fitting regions and the procedure repeated.

Unfortunately, the He~{\sc i} narrow template was found to be too narrow for the corresponding He~{\sc i} component in the mean absorbed spectrum. Given the long time-period of more than ten years between the observation of the reference spectrum and the observations of the absorbed spectra, this disparity is not surprising. Variability of the narrow component of permitted line transitions can occur during such intervals without affecting the validity of the photometric correction factors listed in Table~\ref{tab:obs}, which concern much shorter time-periods over which narrow lines are stable \citep{peterson13}. Following the equivalent procedure for the reference spectrum, we instead fitted the mean spectrum with a new triple-Gaussian model, creating an updated He~{\sc i} narrow template.

\begin{figure}
	\includegraphics[width=\columnwidth]{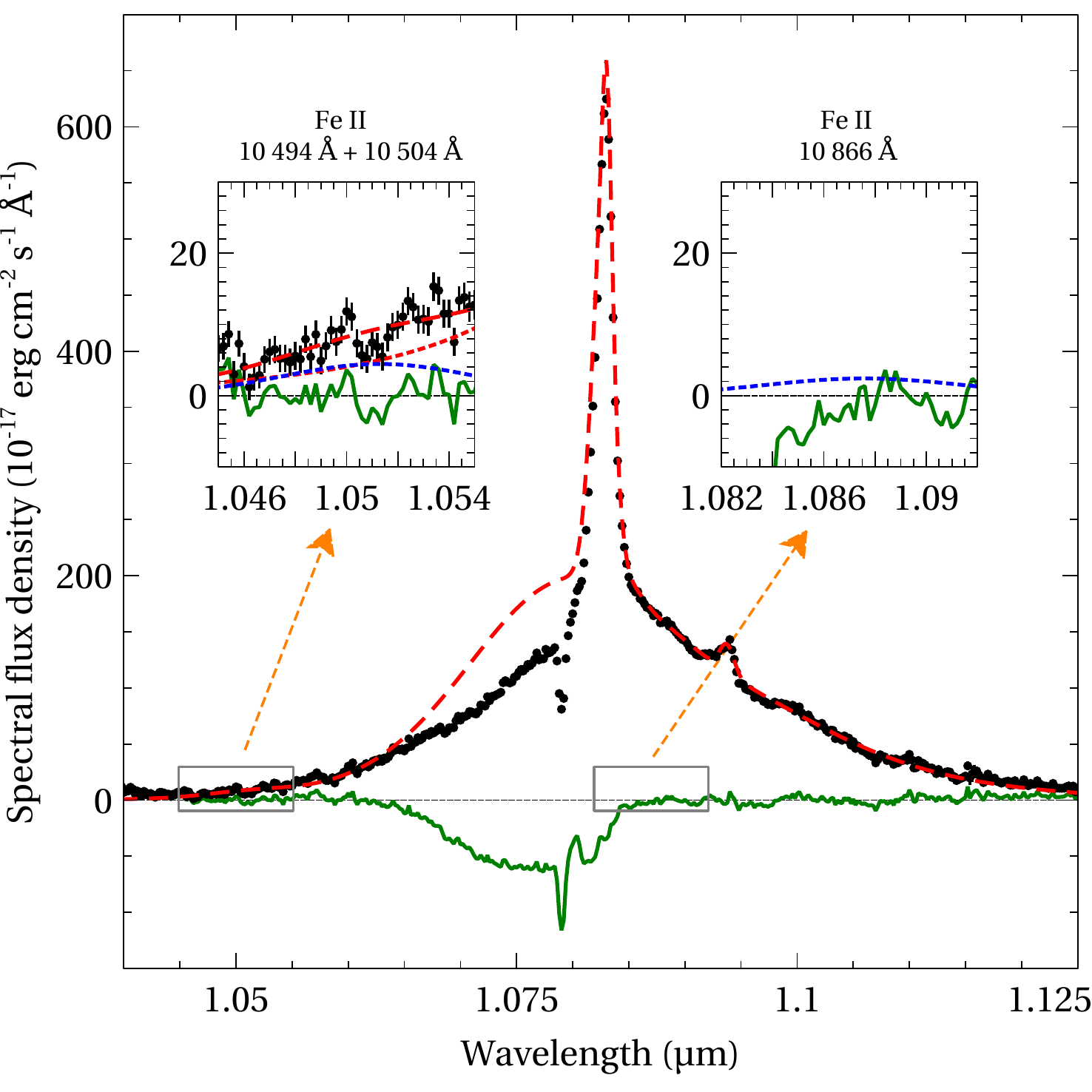}
    \caption{Mean of the eight continuum-subtracted absorbed spectra. Black circles are the observed datapoints and the red dashed-line is the total reconstructed emission. The green line is the difference spectrum (observed minus model), which shows significant absorption on the blue side of the blend. Inset panels are enlargements of the spectral regions enclosed in the grey boxes, which show the Fe~{\sc ii} emission (blue dotted-lines). For the relatively strong 10\,500~\AA{} doublet (left inset panel) the contribution of Fe~{\sc ii} to the total emission is apparent from the red dotted-line, which is the emission model minus the Fe~{\sc ii} model.}
    \label{fig:meanspec}
\end{figure}

Additionally, weak Fe~{\sc ii} emission was included in the mean spectrum fit. Gaussian profiles of mutually locked velocity widths, as well as wavelength ratios locked to the laboratory values, were used to model each of several broad Fe~{\sc ii} lines for this purpose. Flux ratios of Fe~{\sc ii} lines were locked to the theoretical values from \citet{sigut03} which are also listed in \citet{landt08}. The strongest blended Fe~{\sc ii} feature is the doublet at laboratory wavelength $\sim$10\,500~\AA{}, while a weaker feature exists at 10\,866~\AA{}. All components contributing to the total emission fit are listed in Table~\ref{tab:blend}. The complete continuum-subtracted fit, together with the observed and difference spectra, are shown in Fig.~\ref{fig:meanspec}.

\begin{table*}
\begin{center}
\caption{Components of the He~{\sc i} emission line blend}
\begin{tabular}{l l l l}
\hline\hline Component&Lab wavelength&Model used&Mean spectrum FWHM$^\star$\\
 &(\AA{})& &(km~s$^{-1}$)\\
\hline
He~{\sc i} (broad)&10\,833&double Gaussian&6570\\
He~{\sc i} (narrow)&10\,833&triple Gaussian&480\\
Pa$\gamma$ (broad)&10\,941&double Gaussian&8990\\
 & &(fit to broad Pa$\beta$ template)& \\
Pa$\gamma$ (narrow)&10\,941&single Gaussian&410\\
Fe~{\sc ii}&10\,494, 10\,504&double Gaussian&2560, 2560\\
 &10\,866&single Gaussian&2560\\
\hline
\hline    
\end{tabular}
\label{tab:blend}
\end{center}
\parbox[]{12.5cm}{$^\star$Velocities are rounded to the nearest 10~km~s$^{-1}$\\}
\end{table*}

\begin{figure}
	\includegraphics[width=\columnwidth]{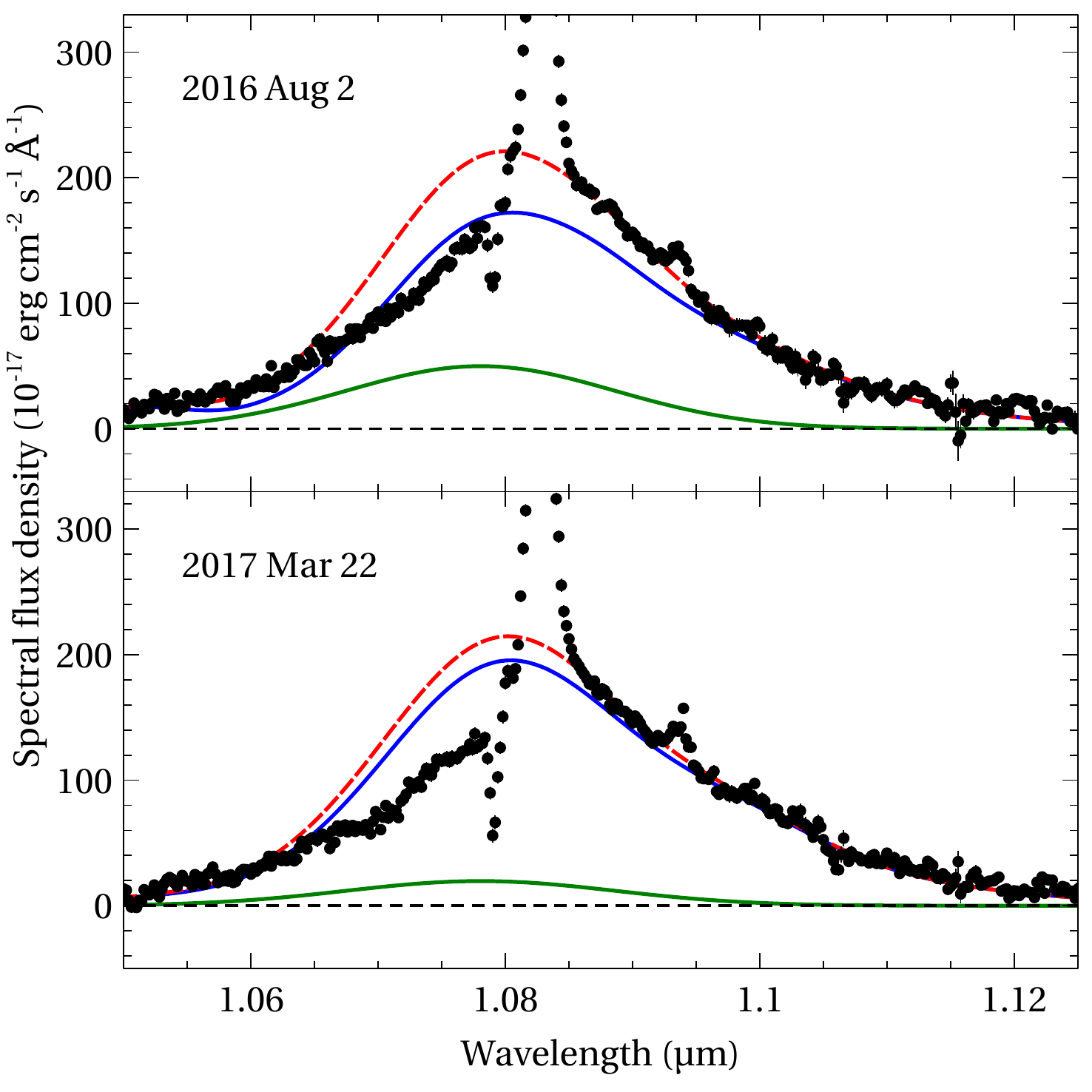}
    \caption{An extra broad component (green) was needed for the fit of these two continuum-subtracted absorbed spectra. The red dashed-line is the total broad blend including the extra component, while the blue line is the same but without the extra component.}
    \label{fig:ep1+7}
\end{figure}

Following this, reconstructions of the unabsorbed, continuum-subtracted spectra of the eight individual absorbed observations were made using the emission components from the fit to the mean spectrum. All Gaussian emission components were locked to their velocity width and wavelength centroid parameters from the mean fit, and the fixed-width He~{\sc i} templates were locked to their respective mean-spectrum centroids. This reduced the free parameters and hence degeneracies in the spectral fitting. The procedure followed then was the same as for the mean spectrum. An exception to this was the creation of a unique Pa$\gamma$ template from Pa$\beta$ for each individual absorbed spectrum, in order to minimize inaccuracies due to variability between observations. This worked well for all observations except for the 1st and 7th observational epochs which required an extra broad Gaussian to satisfactorily fit the blue-wing of the emission line, as illustrated in Fig.~\ref{fig:ep1+7}.

\subsection{Identification of the He~{\sc i} absorbers}

\begin{figure}
	\includegraphics[width=\columnwidth]{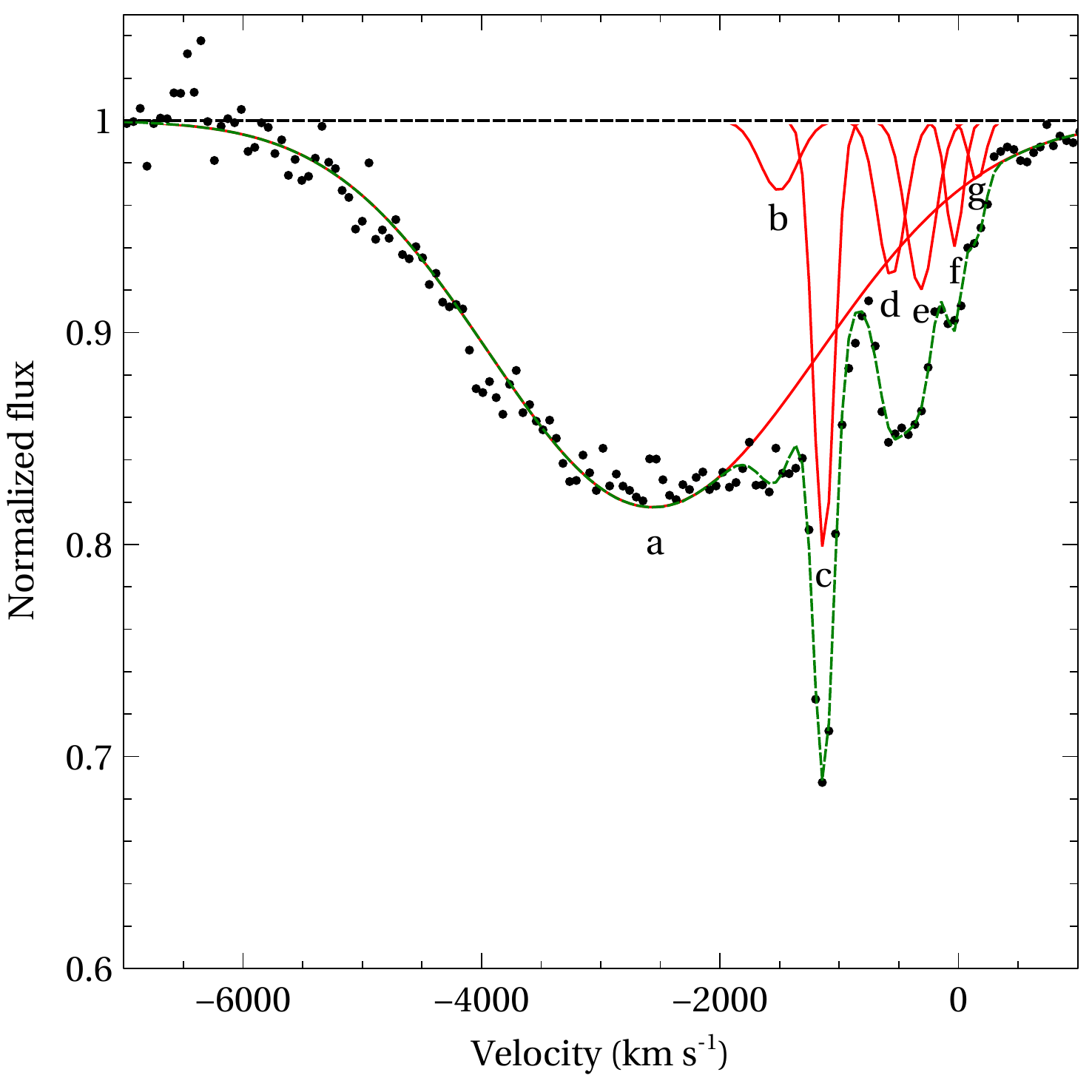}
    \caption{Identified absorbers (solid red lines) in the normalized mean spectrum. Absorbers are labelled at their central velocities. The green dashed-line is the total profile model.}
    \label{fig:meannorm}
\end{figure}

Absorbers were identified in the mean spectrum by adding negative-flux Gaussian components to the reconstruction from Fig.~\ref{fig:meanspec}, in order to fit the observed flux datapoints in the region of absorption. Normalized absorption profiles could then be generated by dividing through by the background illumination. To simplify the process it is assumed that the absorbers are much smaller than both the near-IR dust emission, thought to be located in the torus, and the narrow line emission region, both of which are known to be spatially extended in comparison to intrinsic AGN absorbers. This leaves the emission from the broad line region (BLR) and the accretion disc continuum as the only sources potentially undergoing significant attenuation by these absorbers along our line-of-sight to the AGN. We therefore created a normalized mean absorption spectrum relative to a ``pseudo-continuum'' formed from the sum of disc power-law plus BLR emission. This normalized spectrum, including the identified absorbers labelled \emph{a} through \emph{g}, is shown in Fig.~\ref{fig:meannorm}. From the resulting profile it is apparent that the absorption on the blue-side of the He~{\sc i} blend consists of an overall broad feature (Component \emph{a}) blended with several narrow absorbers.

\begin{figure}
	\includegraphics[width=\columnwidth]{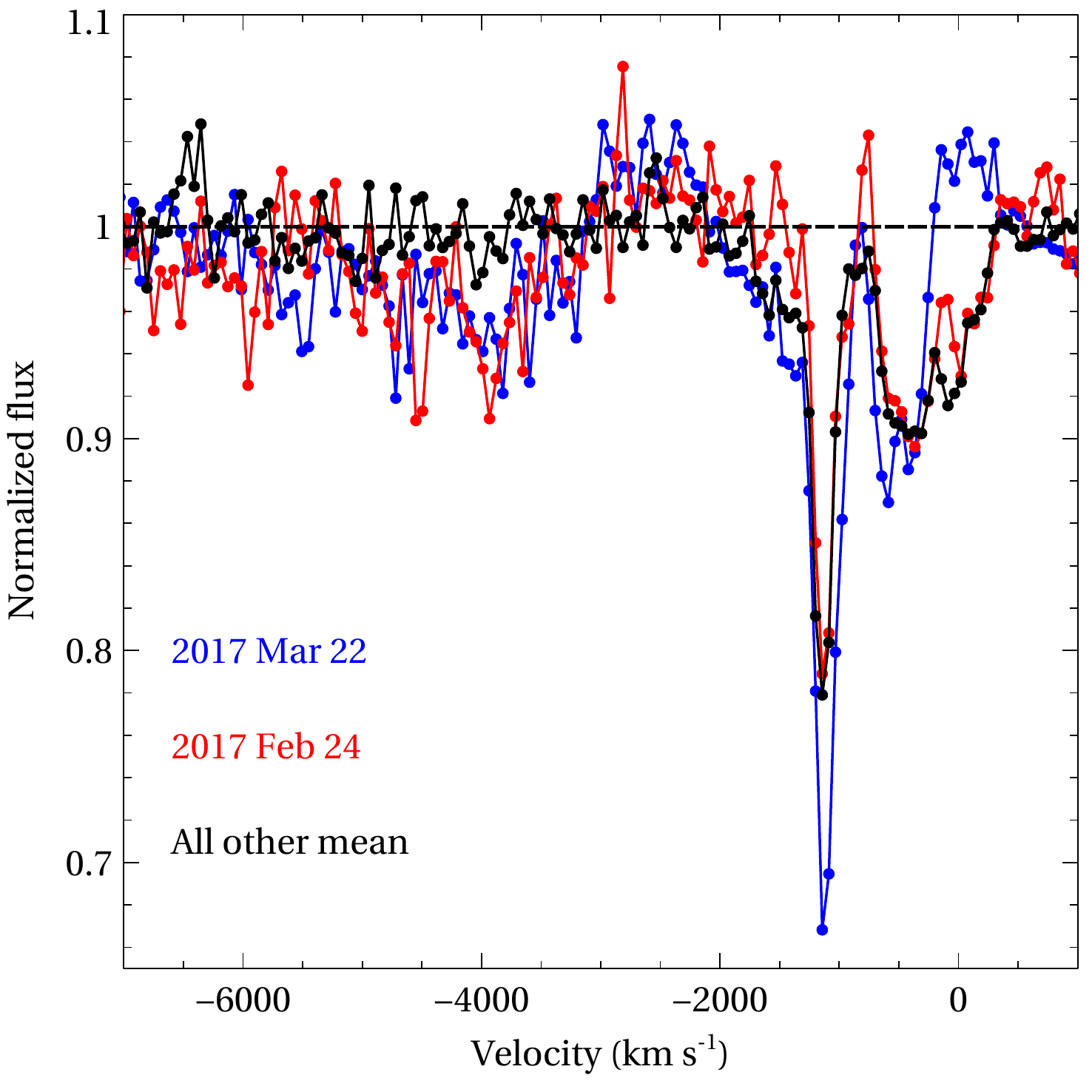}
    \caption{Comparison of the poorly-fitting fifth and seventh-epoch spectra (after normalization to the total emission+absorption profile) with the mean normalized profile for all other spectra.}
    \label{fig:normdiffs}
\end{figure}

\begin{figure}
	\includegraphics[width=\columnwidth]{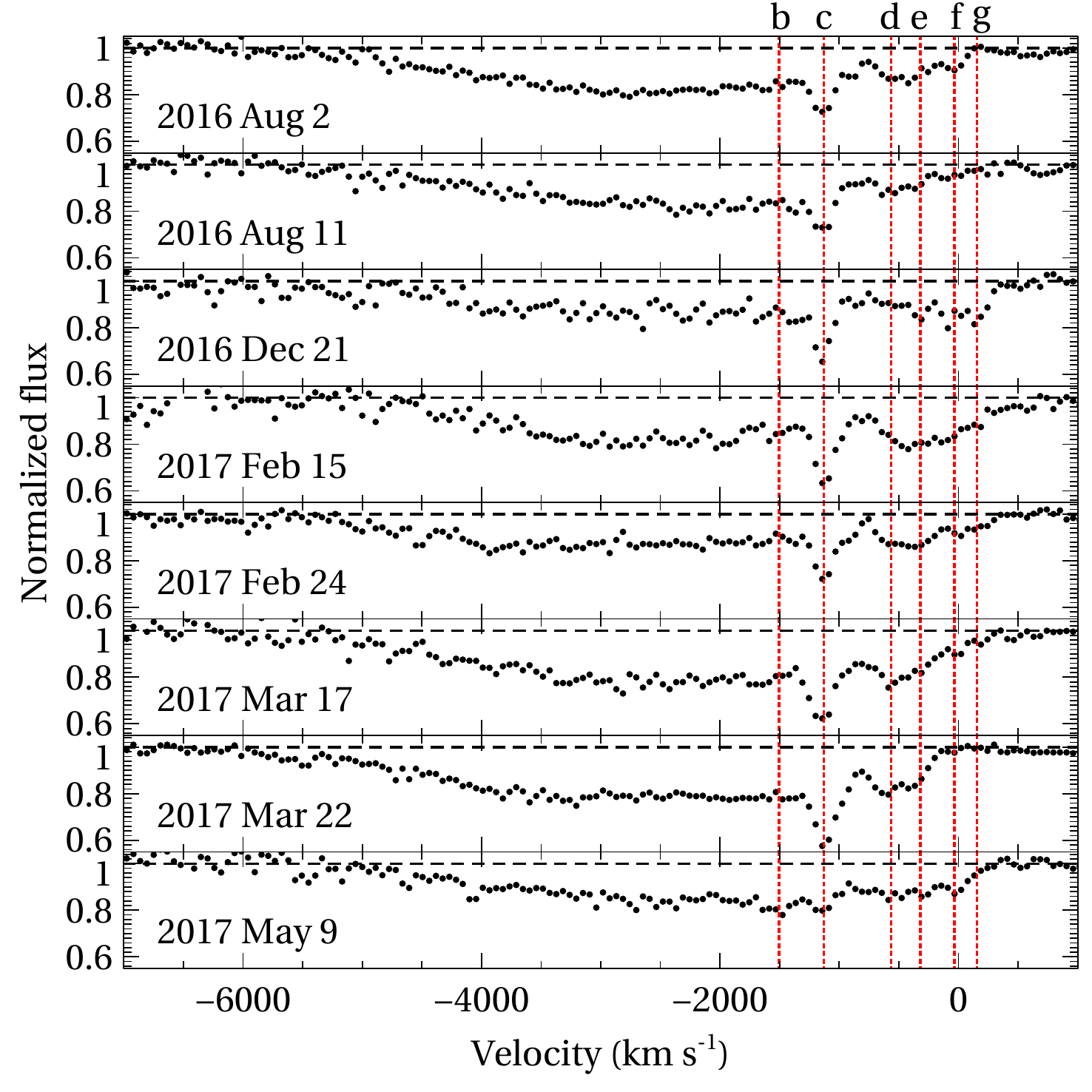}
    \caption{Disc+BLR emission-normalized profiles shown as a function of negative velocity (relative to He~{\sc i} rest-frame wavelength) for all observational epochs. Normalized data points are depicted as black circles. Red vertical dashed-lines show the velocity positions of the narrow absorbers (labelled at top).}
    \label{fig:normspec}
\end{figure}

These absorption components were subtracted from the reconstruction formed for each individual spectrum in a similar fashion. The wavelength position and width were fixed at those found for the mean spectrum, but the strength allowed to vary between each individual spectrum. This total (reconstructed emission minus absorption components) model was found to be an acceptable fit to the observed spectrum in most cases except for those taken at the fifth and seventh observational epochs, where the reduced chi-square ($\chi{}^{2}/\nu{}$) for all pixels in the range $-$7000~km~s$^{-1}$$\leq{}$v$\leq$1000~km~s$^{-1}$ was 2.16 and 2.78 respectively. For all other observations it ranged from 0.94 to 1.41. As depicted in Fig.~\ref{fig:normdiffs}, the observations having poor fits deviate most significantly from the total model at velocities $<$$-$3000~km~s$^{-1}$, where excess absorption is found. In Fig.~\ref{fig:normspec}, emission-normalized profiles for each individual observational epoch are shown, with the positions of the identified narrow absorption components highlighted. All absorber components identified in metastable He~{\sc i} are listed in Table~\ref{tab:normspec}.

Consistent with most previous absorption variability studies, there is no significant shift in individual absorber velocity centroid between observations, however, absorption strengths do vary to a statistically significant degree (Section \ref{variability}). The overall profile and position of Component \emph{a} is similar to the broad absorber seen in recent UV observations and first described in \citet{kaastra14}. Following their analysis, this could indicate that Component \emph{a} is due to the X-ray obscurer that appeared in 2013. \citet{kriss19} associate several broad absorption troughs observed in their 2014 {\sl HST} spectroscopy with this new component of the nuclear structure of NGC~5548, including absorption from ions with a low ionization potential such as H$^+$ and C$^+$ (see their Table 7).

The narrow Component \emph{c} closely matches the velocity centroid and width of Component 1 from \citet{mathur99}, which is labelled UV1 by \citet{arav15}. Narrow components \emph{e} and \emph{f} likewise closely match UV5 and UV6 from the same paper. It is therefore likely that these He~{\sc i} narrow components result from the same outflowing material responsible for the corresponding UV absorbers. Component \emph{d} is intermediate in position to Components UV3 and UV4, and has a broader width. This may be an artefact of the lower resolution of our observations, which record the two as blended into a single apparent absorber. This seems especially likely given the presence of a Si~{\sc iv} absorption feature seen in Figure~9 of \citet{kriss19} centered at approximately 600~km~s$^{-1}$. Like metastable He~{\sc i}, Si~{\sc iv} exists at moderate ionization conditions, meaning absorption in one likely shows up in the other. As discussed by \citet{arav15}, UV1 and UV3 also have WA counterparts in the X-ray data of \citet{kaastra14}. Narrow component $\emph{b}$ does not have any obvious counterpart from the UV data of previous studies. But narrow component \emph{g}, the only inflowing (rather than outflowing) absorber in our data, is most likely related to Component~6 of \citet{mathur99}, who measured a (positive) velocity of $\sim 250$~km~s$^{-1}$. This component was also seen later by \citet{crenshaw99} in Ly$\alpha$ only, when it had a velocity of $\sim 90$~km~s$^{-1}$. 

\begin{table}
\begin{center}
\caption{List of identified He~{\sc i} absorber properties and corresponding UV component from \citet{arav15}.}
\begin{tabular}{l l l l}
\hline\hline Component&Centroid velocity$^\star$&Width$^\star$&UV component\\
 &(km~s$^{-1}$)&(km~s$^{-1}$)\\
\hline
a&$-$2360&3090&broad absorber\\
b&$-$1510&370&none\\
c&$-$1130&210&UV1\\
d&$-$560&270&UV3+UV4\\
e&$-$320&290&UV5\\
f&$-$30&170&UV6\\
g&$\phantom{-}$150&160&none\\
\hline
\hline    
\end{tabular}
\label{tab:normspec}
\end{center}
$^\star$ Velocities are rounded to the nearest 10~km~s$^{-1}$
\end{table}

\subsection{Variability measurements} \label{variability}

There are several different methods available for measuring the variability of individual absorption components. A simple method is to measure changes in the total flux within a component, while an alternative is to measure it with respect to the background emission source using the equivalent width (EW). In both cases it is important to estimate errors in order to quantify the significance of any variability. It is assumed that the contribution to the total errors by the absorption and emission models is dominated by the fractional uncertainty in their strength normalization rather than their shape. The estimated value of this fractional error for each absorption model was obtained by incrementally changing the strength of each by 0.01 percent, both in the increasing and decreasing direction, until the criterion for a 1$\sigma{}$ change for one interesting parameter ($\Delta{}\chi{}^{2}>$1) was reached. 

The wavelength range used for the fractional error calculations for the continuum components were those previously defined for the S/N calculation, while for the absorption models the velocity range is the same as that over which $\chi{}^{2}/\nu{}$ was calculated. Uncertainties in the narrow and broad emission models are calculated from the blue (unabsorbed) side of the respective component in the same way. For simplicity but while remaining conservative, symmetrical errors are assumed, with the larger value of the two directions adopted. For the final 1$\sigma$ uncertainty in both the EW and total flux, the fractional model errors are propagated together with the per pixel observed errors to determine the total error.

\begin{figure}
	\includegraphics[width=\columnwidth]{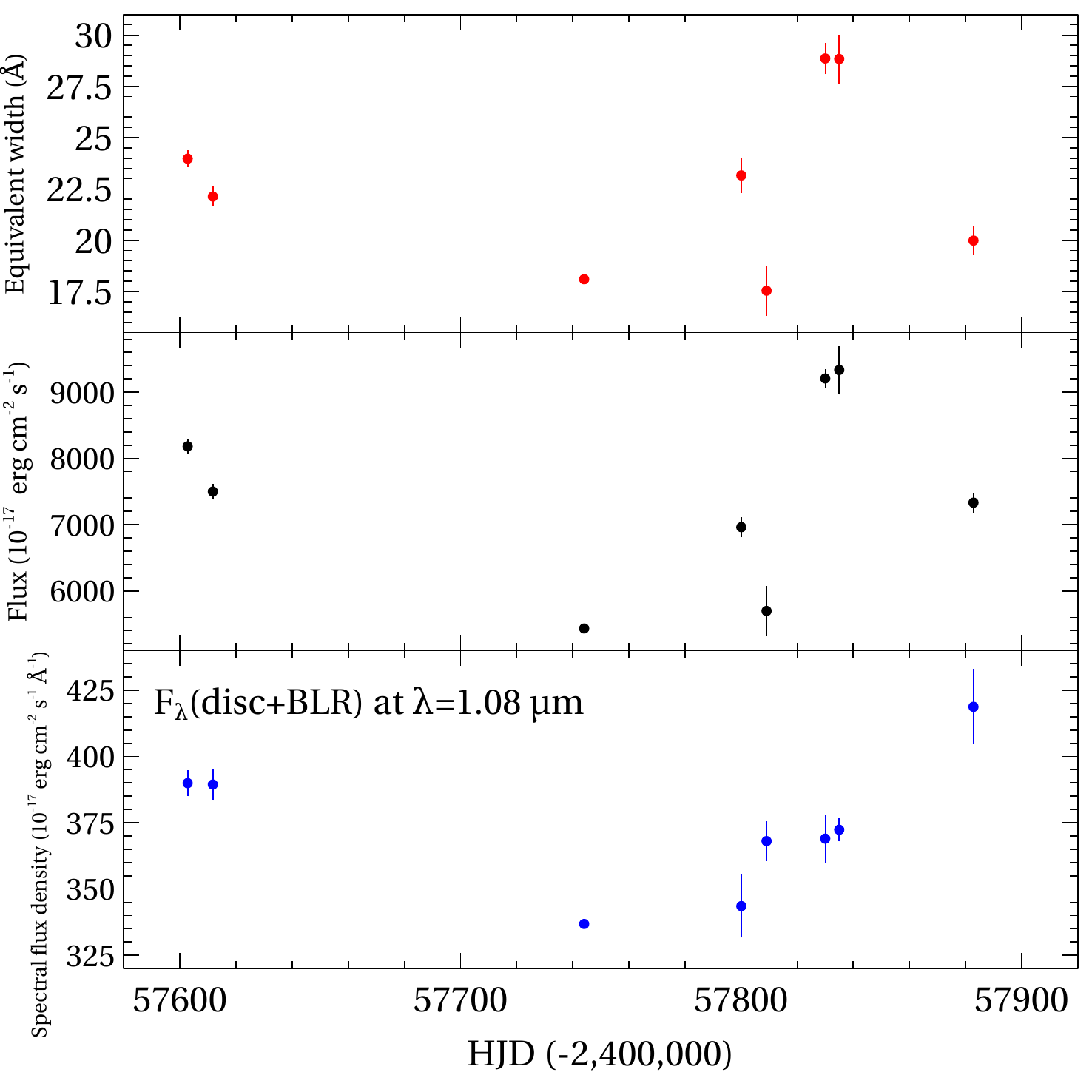}
    \caption{\emph{Top panel}: Red points mark the EW measurements (relative to disc+BLR emission) of the broad absorption component as a function of time. \emph{Middle panel}: Black points mark the total absorbed flux measurements of the broad absorption component as a function of time. \emph{Bottom panel}: Blue points show the pseudo-continuum variation as a function of time.}
    \label{fig:broadvar}
\end{figure}

The broad absorber must cover a portion of the BLR along our line-of-sight, given the depths of the UV troughs recorded in \citet{kaastra14}. Since the reverberation-measured size of the hydrogen BLR is $\sim 4$~light-days \citep{pei17}, the disc emission region at 1.08$\mu{}$m, under the assumption of an optically thick, geometrically thin disc \citep{shakura73}, likely occurs at a similar distance from the ionizing source. We therefore make the simplifying assumption that, as in the UV, the broad absorber ``sees'' only the disc+BLR pseudo-continuum along our line-of-sight. The variability behaviour of the broad absorption component and the pseudo-continuum across the observation period are illustrated in Fig.~\ref{fig:broadvar}, while the absorber properties are tabulated in Table \ref{tab:broadfluxes}. Cases of significant variability (at least 2$\sigma{}$) are seen between many pairs of observations. Both the EW measurement and the total flux show the same variability pattern over time.

\begin{table}
\caption{Fluxes, equivalent widths and ionic column densities for component $a$.}
\begin{tabular}{l l l l}
\hline\hline Observational & Flux & Equivalent & Column \\
epoch number & (10$^{-17}$~erg~cm$^{-2}$~s$^{-1}$) & width & density \\
 && (\AA) & (10$^{13}$~cm$^{-2}$) \\
\hline
1&8180$\pm$110&24.0$\pm$0.42&4.65$\pm$0.09\\
2&7500$\pm$121&22.1$\pm$0.48&4.27$\pm$0.10\\
3&5430$\pm$153&18.1$\pm$0.67&3.44$\pm$0.14\\
4&6960$\pm$151&23.2$\pm$0.85&4.48$\pm$0.18\\
5&5700$\pm$381&17.5$\pm$1.21&3.33$\pm$0.24\\
6&9210$\pm$143&28.9$\pm$0.76&5.69$\pm$0.16\\
7&9340$\pm$373&28.8$\pm$1.19&5.68$\pm$0.26\\
8&7330$\pm$150&20.0$\pm$0.71&3.82$\pm$0.15\\ 
\hline
\hline    
\end{tabular}
\label{tab:broadfluxes}
\end{table}

\begin{figure}
	\includegraphics[width=\columnwidth]{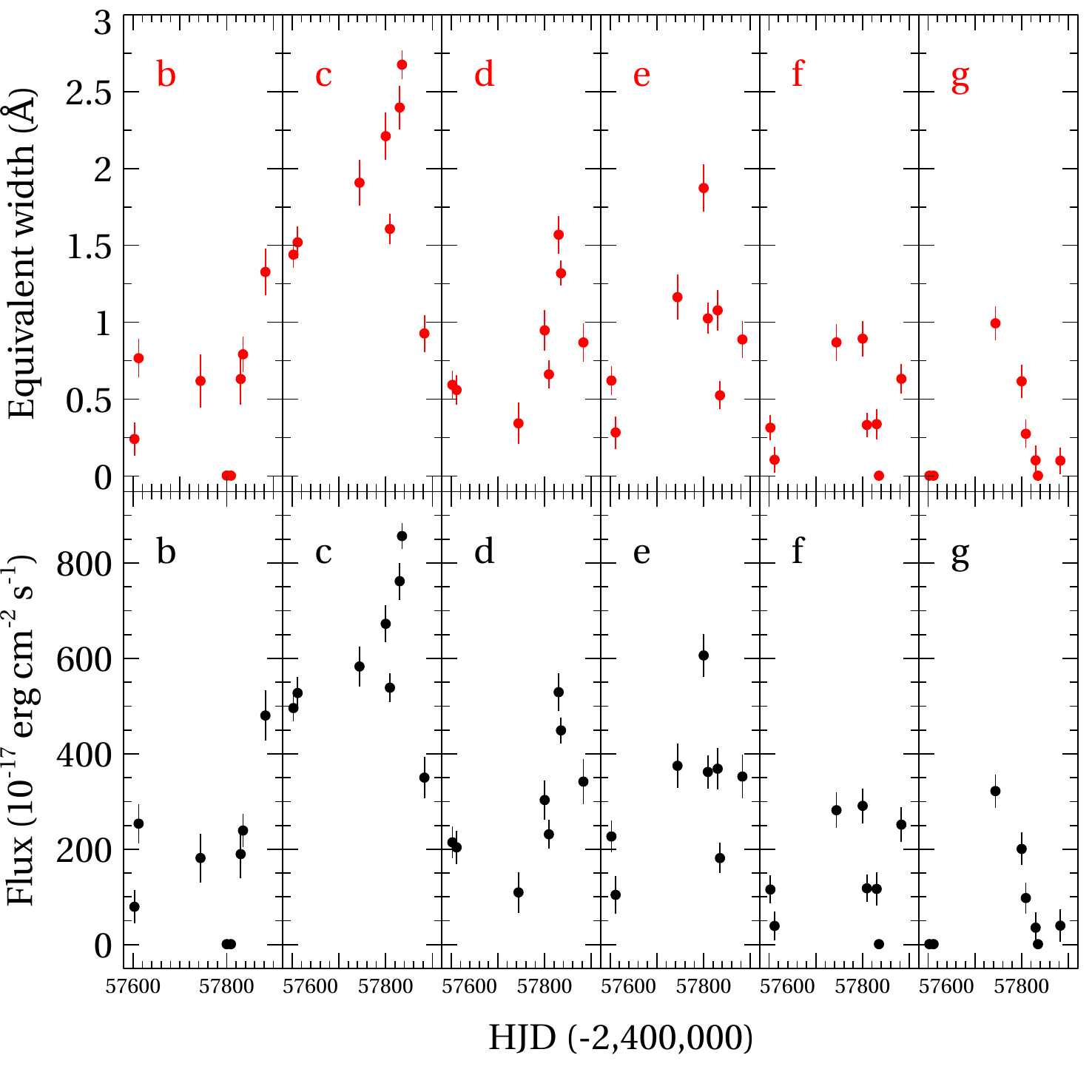}
    \caption{\emph{Top panels}: Red points mark the EW measurements (relative to disc+BLR emission after passing through the broad absorber) of the narrow absorption components as a function of time. \emph{Bottom panels}: Black points mark the total absorbed flux measurements of the narrow absorption components as a function of time. Component labels are positioned top-left.}
    \label{fig:narrowvar}
\end{figure}

The view of the disc+BLR pseudo-continuum along the line-of-sight from the warm X-ray absorber components, which are responsible for at least some of the narrow UV absorbers, is believed to be obstructed by the obscurer \citep{kaastra14}. Therefore, EW measurements for the narrow components are calculated from a new pseudo-continuum, using not only disc+BLR as before, but also including attenuation of this continuum by the broad absorber. The variability of the narrow components is shown in Fig.~\ref{fig:narrowvar}, with their values tabulated in Table~\ref{tab:narrowfluxes}. Similarly as for the broad absorber, significant variability is seen in all narrow absorption components. Errors are only calculated for a particular narrow absorber for those observations where its strength is non-zero.

\begin{table*}
\begin{center}
\caption{Fluxes and equivalent widths for components $b-g$.}
\begin{tabular}{l l l l l l l}
\hline\hline Observational & \multicolumn{2}{c}{component $b$} & \multicolumn{2}{c}{component $c$} & \multicolumn{2}{c}{component $d$} \\
epoch number & Flux & Equivalent width & Flux & Equivalent width & Flux & Equivalent width \\
&(10$^{-17}$~erg~cm$^{-2}$~s$^{-1}$) & (\AA) & (10$^{-17}$~erg~cm$^{-2}$~s$^{-1}$) & (\AA) & (10$^{-17}$~erg~cm$^{-2}$~s$^{-1}$) & (\AA) \\
\hline
1&79.4$\pm$35.3&0.24$\pm$0.11&496$\pm$28.2&1.44$\pm$0.08&214$\pm$33.7&0.59$\pm$0.09\\
2&254$\pm$41.0&0.77$\pm$0.13&528$\pm$34.3&1.52$\pm$0.10&204$\pm$34.8&0.56$\pm$0.10\\
3&182$\pm$50.9&0.62$\pm$0.17&583$\pm$41.5&1.91$\pm$0.15&110$\pm$42.4&0.34$\pm$0.13\\
4&0&0&673$\pm$38.9&2.21$\pm$0.15&303$\pm$41.1&0.95$\pm$0.13\\
5&0&0&539$\pm$30.4&1.61$\pm$0.10&231$\pm$30.7&0.66$\pm$0.09\\
6&190$\pm$50.2&0.63$\pm$0.17&762$\pm$38.9&2.40$\pm$0.14&529$\pm$39.1&1.57$\pm$0.12\\
7&239$\pm$35.1&0.792$\pm$0.12&857$\pm$26.9&2.67$\pm$0.09&449$\pm$27.2&1.32$\pm$0.08\\
8&480$\pm$52.4&1.33$\pm$0.15&350$\pm$43.5&0.93$\pm$0.12&342$\pm$47.7&0.87$\pm$0.13\\ 
\hline 
Observational & \multicolumn{2}{c}{component $e$} & \multicolumn{2}{c}{component $f$} & \multicolumn{2}{c}{component $g$} \\
epoch number & Flux & Equivalent width & Flux & Equivalent width & Flux & Equivalent width \\
&(10$^{-17}$~erg~cm$^{-2}$~s$^{-1}$) & (\AA) & (10$^{-17}$~erg~cm$^{-2}$~s$^{-1}$) & (\AA) & (10$^{-17}$~erg~cm$^{-2}$~s$^{-1}$) & (\AA) \\
\hline
1&227$\pm$33.4&0.62$\pm$0.09&116$\pm$29.7&0.31$\pm$0.08&0&0\\
2&104$\pm$38.9&0.28$\pm$0.11&39.2$\pm$30.5&0.11$\pm$0.08&0&0\\
3&375$\pm$46.1&1.16$\pm$0.15&282$\pm$37.8&0.87$\pm$0.12&322$\pm$35.1&0.99$\pm$0.11\\
4&606$\pm$45.1&1.87$\pm$0.16&291$\pm$36.4&0.89$\pm$0.12&201$\pm$34.1&0.62$\pm$0.11\\
5&362$\pm$34.5&1.03$\pm$0.10&118$\pm$28.6&0.33$\pm$0.08&97.7$\pm$32.3&0.28$\pm$0.09\\
6&369$\pm$43.5&1.08$\pm$0.13&117$\pm$34.3&0.34$\pm$0.10&35.6$\pm$32.8&0.10$\pm$0.09\\
7&181$\pm$31.7&0.52$\pm$0.09&0&0&0&0\\
8&352$\pm$46.2&0.89$\pm$0.12&252$\pm$36.7&0.63$\pm$0.09&39.8$\pm$34.4&0.10$\pm$0.09\\ 
\hline
\hline    
\end{tabular}
\label{tab:narrowfluxes}
\end{center}
\end{table*}

\section{Discussion}

\subsection{The broad absorption component} \label{obscurer}

The broad absorption component $a$ appears similar to the UV absorption described in \citet{kaastra14}, which is likely responsible for the X-ray obscuration first seen in 2013. The absence of broad absorption in UV spectra before then is consistent with the absence of broad He~{\sc i} absorption in the June 2006 spectrum, supporting the interpretation that both have a common origin. The broad UV absorption lines seen in \citet{kaastra14} were observed at saturation point, as deduced from the almost 1:1 red-blue Si~{\sc iv} component ratio, but did not reach zero intensity. Such a situation implies partial coverage of the emission source. For a coordinated change in strength across a broad absorption line between observations, the most likely explanation is a change in the ionization state of the absorbing material in response to variations in the incident ionizing continuum; as the ionization parameter increases, the equivalent width of the absorption line decreases. An alternative explanation is that the density of the obscurer varied, as investigated in detail by \citet{dehghanian19b}, with a higher density leading to a lowered absorption line equivalent width (see their Fig. 4). 

\begin{figure}
	\includegraphics[width=\columnwidth]{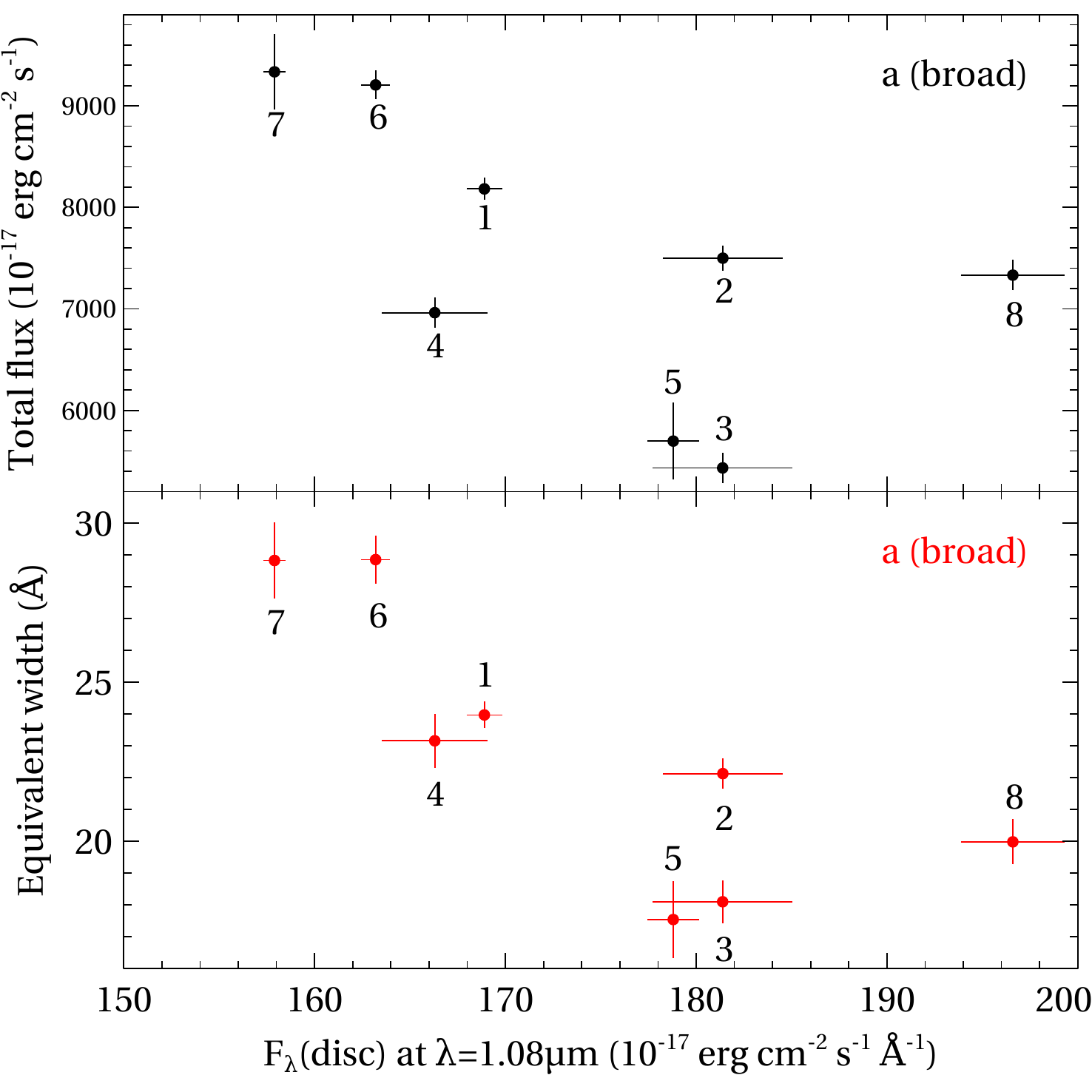}
    \caption{\emph{Top panel}: Relationship between total flux of component $a$ and NIR continuum strength \emph{Bottom panel}: Relationship between EW of component $a$ and NIR continuum strength. The numbers represent the time order of the observations from the earliest to the most recent.}
    \label{fig:broadvdisk}
\end{figure}

There is a significant anti-correlation between the strength of component $a$ and the strength of the disc continuum at 1.08~$\mu$m, as shown in Fig.~\ref{fig:broadvdisk}. This is visible in both the EW and total flux measures of the absorber strength. If the disc NIR continuum at this wavelength is indicative of changes in the ionizing EUV spectral region and soft X-rays, it shows that ionization changes in the obscurer are likely influencing the strength of the absorption line. Since a saturated absorption line cannot undergo a strength increase in response to changes in ionization state, these results suggest that component $a$ is not saturated in He~{\sc i} $\lambda$10\,833 for most, or possibly all, of these observations. But Fig.~\ref{fig:broadvdisk} also shows that a significant drop in equivalent width can occur between observational epochs although the disc emission has barely changed. Therefore, it is likely that the density of the obscurer also varies.  

In the presence of an unsaturated absorption line, it is possible to determine the ionic column density and ionization parameter of the gas. We have calculated the apparent metastable helium column density at each observational epoch from the component $a$ models. It should be noted that these are lower limits to the true value if the profile {\it is} saturated (and only partially covers the emission source due to non-zero emission in the line profile). The calculation was performed by directly integrating the component $a$ absorption profile, normalized to the disc+BLR pseudo-continuum, according to the method of \citet{savage91} as shown in Equation~\ref{eqn:cdencalc}

\begin{equation}
\label{eqn:cdencalc}
N_{He}=\frac{m_{e}c}{\pi{}e^{2}f\lambda{}}\int \tau{}\left(v\right)dv
\end{equation}

\noindent where ${N_{He}}$ is the apparent metastable He~{\sc i} column density, ${m_{e}}$ is the electron mass, $c$ is the vacuum speed-of-light, $f$ is the oscillator-strength of the transition, $v$ is velocity and $\tau(v)$ is the apparent optical depth at velocity $v$. The optical depth can be obtained using $\tau(v)$=$ln(1/f_{n}(v))$, where $f_{n}(v)$ is normalized flux at velocity $v$. Calculated values are listed in Table~\ref{tab:broadfluxes}.

\subsubsection{Photoionization simulations}

The X-ray obscurer described in \citet{cappi16} has two components, one is neutral and the other moderately ionized. The two most recent observations listed in Table 4 of \citet{cappi16} are labelled \emph{M13N} and \emph{M14}, observed on 2013 Dec 20-21 and 2014 Feb 04-05, respectively. We performed a test with the \emph{Cloudy} software, version C17.01 \citep{ferland17}, to check if the properties of the moderately ionized component of the X-ray obscurer reported in \citet{cappi16} were consistent with the component $a$ model. This was achievable since \citet{cappi16} were able to determine an ionization parameter for the moderately ionized component from X-ray analysis, which was not possible for the neutral component. However, it should be noted that the neutral component could also contain metastable He~{\sc i} and thus contribute to component $a$.

The continuum input to \emph{Cloudy} was set to the spectral energy distribution (SED) given by the \textit{Table SED ``NGC 5548''} command, which adopts the absorption-corrected continuum for NGC~5548 shown in Figure~10 of \citet{mehdipour15}. We assumed a number density of log($n_{H}$ / cm$^{-3}$)=10 for the broad absorber, which is within the range allowed for BLR \citep{osterbrock06,popovic03}. Using the listed hydrogen column density and ionization parameter values, the corresponding metastable He~{\sc i} column densities were calculated to be 1.1$\times{}$10$^{15}$~cm$^{-2}$ for \emph{M13N} and 3.5$\times{}$10$^{12}$~cm$^{-2}$ for \emph{M14}. The values listed in Table~\ref{tab:broadfluxes} are intermediate to these two values, so there is no contradiction with the \citet{cappi16} data for the moderately ionized component, although it should be noted that the two He~{\sc i} column densities derived from that data show a large change in a very short time-frame.

\citet{kriss19} presented photoionization model constraints on the obscurer using a set of absorption lines from eight different ions identified in their high-spectral resolution HST COS UV spectra from the 2014 campaign. Some of these lines were saturated and did not converge to a single solution. However, the weaker, unsaturated absorption lines constrained the moderately ionized component of the obscurer (i.e. the component with the lower hydrogen column density as derived from X-ray observations) to a ionization parameter of log~$\xi=0.8-0.95$ and the neutral component of the obscurer to an even higher ionization parameter of log~$\xi=1.5-1.6$. 

\begin{figure}
	\includegraphics[width=\columnwidth]{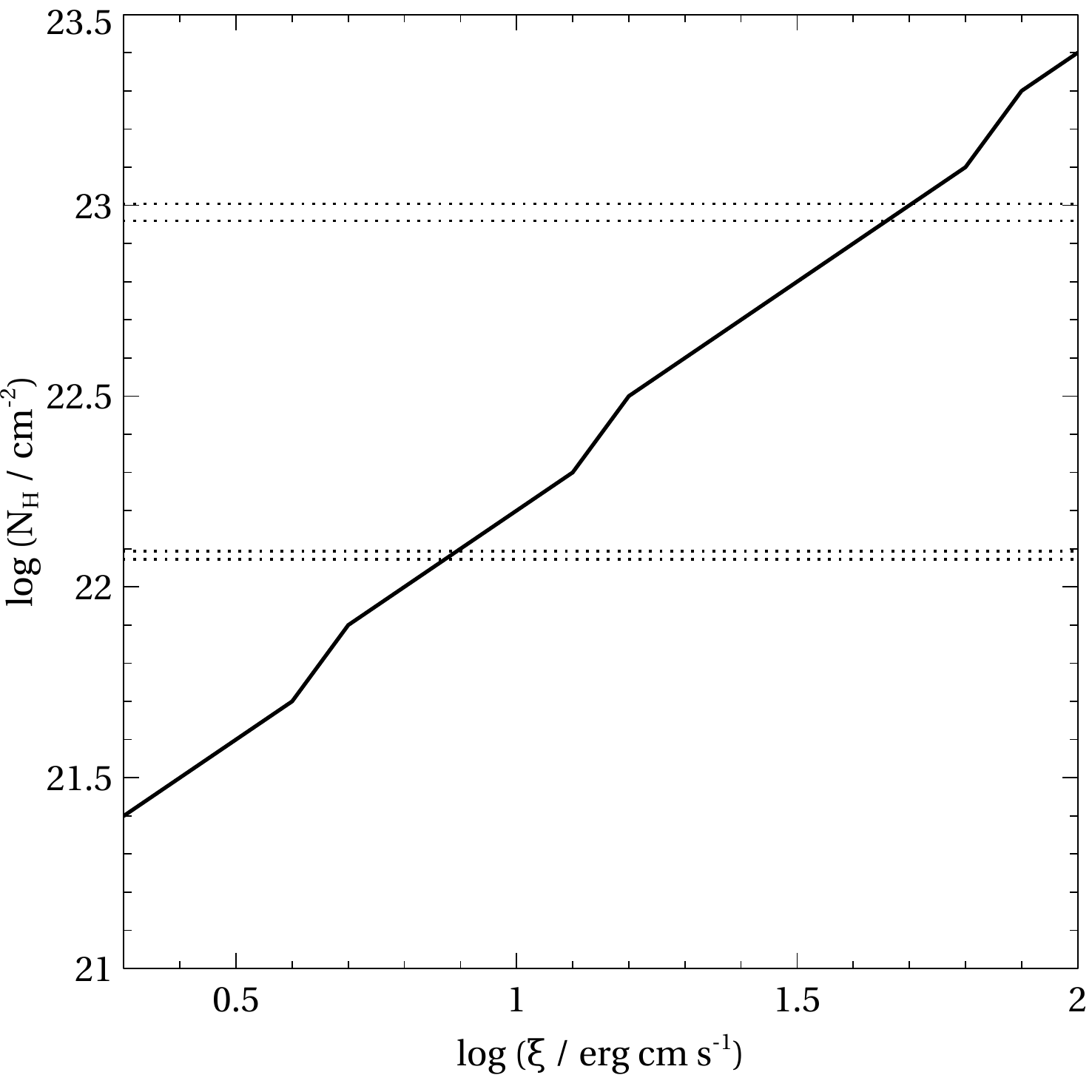}
    \caption{Photoionization model constraints on the obscurer in NGC~5548. Dotted lines give the range of total column density for the two obscurer components as determined from the 2013/14 X-ray observations. The solid line shows the run of column density with ionization parameter for metastable He~{\sc i}.}
    \label{fig:ionparameter}
\end{figure}

Using the same SED and number density as before, we have converted the ionic column densities to hydrogen column densities for a range of ionization parameters using the \emph{Cloudy} software. This was done by calculating the metastable He~{\sc i} column density for all points in a grid spanning 0.3$\leq$log~$\xi$$\leq$2.0 and 21.0$\leq$log~$N_{H}$$\leq$23.5 of step size 0.1 in both dimensions. Then, for each value of log~$\xi$, the log$N_{H}$ gridpoint resulting in the value closest to the average of all helium column densities listed in Table~4 (4.4$\times$10$^{13}$~cm$^{-2}$) is recorded. The path of these recorded points through the grid space is shown in Fig.~\ref{fig:ionparameter}. Similar to \citet{kriss19}, we obtain ionization parameters of log~$\xi \sim 0.9$ and log~$\xi \sim 1.7$ for the obscurer components with measured hydrogen column densities in 2013/14 of log~$N_{\rm H} \sim 22.1$~cm$^{-2}$ and log~$N_{\rm H} \sim 23.0$~cm$^{-2}$, respectively.

\subsection{The narrow absorption components}

High-ionization narrow absorption lines are known to be present over decades from previous spectral observations of NGC~5548. Previous studies show they were accompanied by moderate and low ionization lines at the same outflow velocities after the appearance of the X-ray obscurer. As noted before, metastable He~{\sc i} is effectively a moderately ionized species, existing over ionizing conditions similar to those of Si~{\sc iv}. The absence of either narrow or broad metastable helium absorption lines in 2006 June, followed by their presence in 2016/17, is therefore consistent with the appearance of other low to moderate-ionization narrow absorption lines after the first detection of the X-ray obscurer. From the equivalent width and total flux analysis, it is clear that the narrow components show significant variability in strength across the whole time span of the observations. 

\begin{figure}
	\includegraphics[width=\columnwidth]{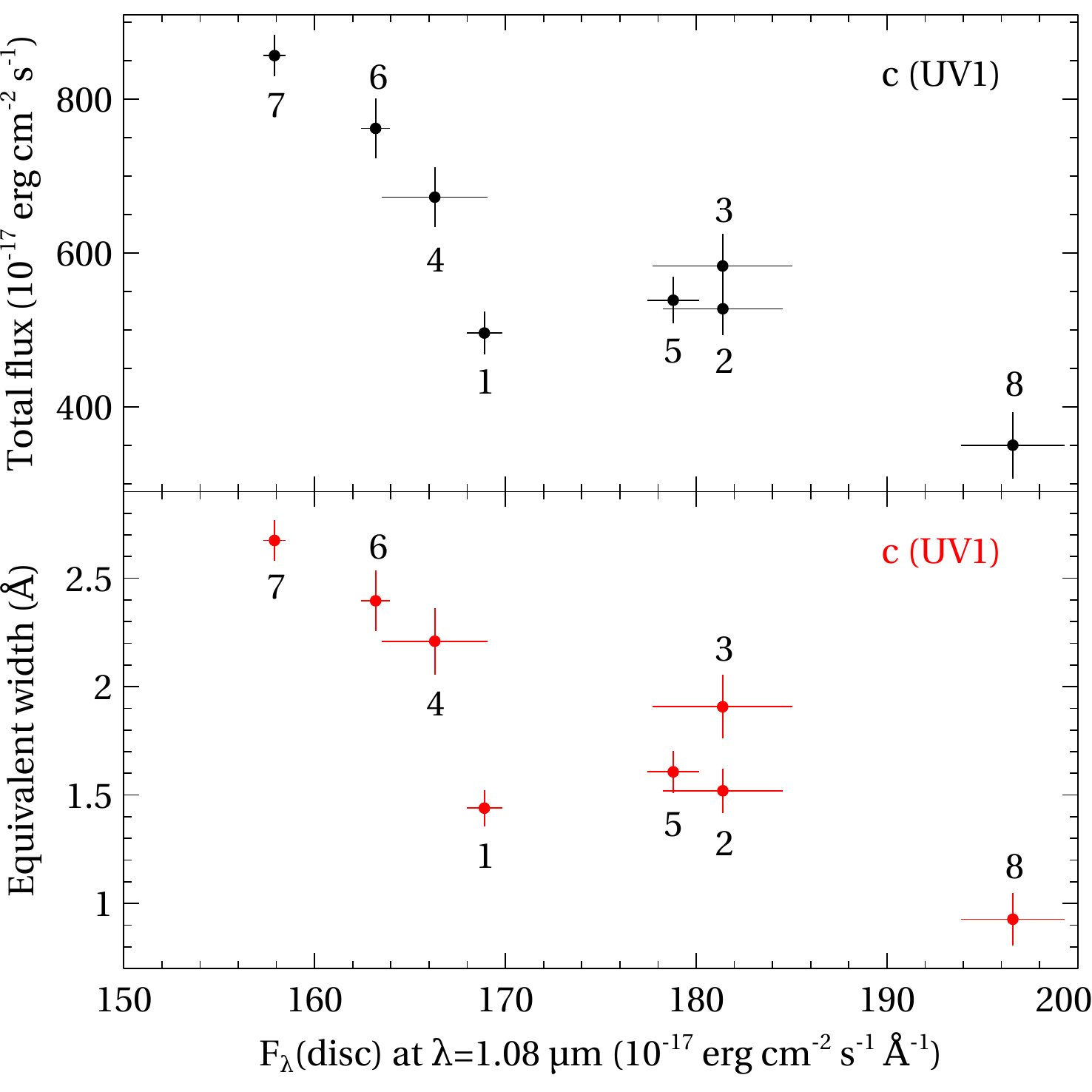}
    \caption{\emph{Top panel}: Relationship between total flux of component $c$ (UV1) and NIR continuum strength \emph{Bottom panel}: Relationship between EW of component $c$ (UV1) and NIR continuum strength. Numbers represent the time order of the observations from the earliest to the most recent.}
    \label{fig:uv1abs}
\end{figure}

A strong anti-correlation is found between the strength of the disc NIR continuum and the strength of component $c$, which we associate with the UV absorber ``UV1''. This is  illustrated in Fig.~\ref{fig:uv1abs} and suggests that a similar ionization process, as was the case for the broad component, is producing changes in the strength of this absorption feature. No other narrow absorption component shows a similar strong relationship with the changes in the disc NIR continuum, which suggests variability in these components is not related to ionization changes, assuming that the disc NIR continuum is a proxy for the ionizing EUV continuum and soft X-rays.

A possibility for the origin of component $b$, which does not appear in the UV spectra of previous studies, is that it absorbs the dust continuum rather than the disc+BLR. Such lines would not show up in the UV as the dust does not emit at those wavelengths. Such absorbing material would be physically distant from the UV absorbers since the line-of-sight to the dust continuum, at the periphery of the AGN emission region, is very different from that for the central disc emission region. Its origin could in fact be the dusty torus itself; e.g. colder material located further away and emitting mainly at mid- to far-infrared wavelengths.  This possibility cannot be ruled out by excessive flux in component $b$, as the absorption line model is weaker than the dust emission model across the entire wavelength span of the component. If this component indeed originates in the dusty torus itself, it would explain why it is the narrow absorber with the largest (negative) velocity and width.

As was seen in Fig.~\ref{fig:normdiffs}, there is excess absorption present at the blue-end of the total absorption profile in the spectra taken at both 2017 Feb 24 and 2017 Mar 22. From visual inspection it appears that these features are a collection of narrow absorbers. The observation interposed between these two, on 2017 Mar 17, does not show strong evidence of these features. It is likely that they were lost in the noise as a result of the low quality of this spectrum (see continuum S/N in Table~\ref{tab:obs}). The excess absorption therefore appeared at some point between 2017 Feb 15 and Feb 24 and disappeared before May 9. The continuum behaviour in this period is not unusual, supporting the concept that this absorbing material does not persist over the long-term along our line-of-sight to the continuum source. In that case it would need to exist in the correct ionization state to be seen in metastable He~{\sc i} only during this period. For an emission region a few light days across, an eclipsing cloud moving across the line-of-sight at a speed on the order of 10\,000~km~s$^{-1}$ is a possibility on this timescale.

\subsection{The influence of the obscurer on the warm absorber}

The 2014 intensive HST~spectroscopic reverberation mapping programme on NGC~5548 yielded a surprising result; some of the broad emission lines {\it and} narrow absorption lines did not respond to changes in the EUV continuum flux for some period of the campaign \citep{goad16, pei17, kriss19}. The term ``line holiday'' is now used to refer to this event. Recently, \citet{dehghanian19a,dehghanian19b} proposed a physical cycle that can explain the ``line holiday''. Pointing out that helium is an excellent opacity source for soft X-rays and at the same time the radiation field resulting from recombining helium is an important ionization source for some lines, they were able to explain the influence the obscurer has on the physical state not only of the BLR but also on that of the warm absorber. In their model, the more soft X-rays the obscurer absorbs, the less are available for the warm absorber. This deficiency in soft X-rays then leads to less helium recombination within the warm absorber gas, which in turn leads to weaker lines from ions that are ionized by energy from this radiation field. Therefore, only certain line species are expected to show a ``line holiday'', whereas lines that are ionized directly by the EUV radiation to which the obscurer is transparent will not. 

With our data we can now test this physical cycle since we directly observe helium from both the obscurer and the warm absorber. We assume that the metastable He~{\sc i} absorption line flux from our broad component $a$ relates to the strength of the soft X-ray absorption by the obscurer in the sense that the absorption line flux is smaller the more soft X-rays are absorbed, and so the higher the ionization parameter (for an unchanged density; see Fig.~\ref{fig:broadvdisk}). For the strongest of the warm absorbers, our component $c$ (referred to as Component~1 in the study of \citet{dehghanian19a}), we expect that the influence of the obscurer on it is manifested as an increase in narrow absorption line flux when the broad absorption line flux decreases and vice versa. In effect, the obscurer and the warm absorber gas compete for the soft X-rays, which is expected to create an interdependence detectable in He~{\sc i} absorption.

\begin{figure}
	\includegraphics[width=\columnwidth]{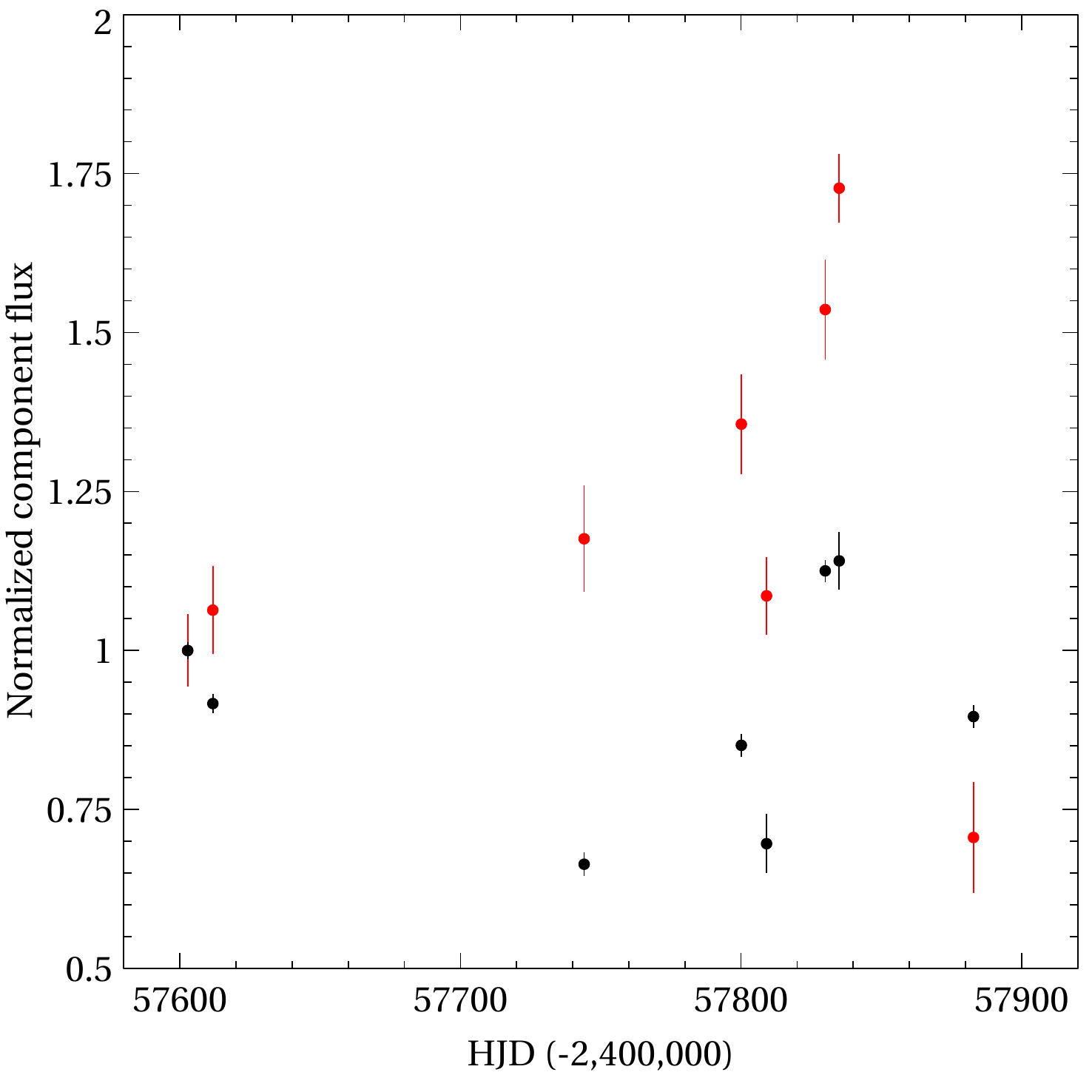}
    \caption{Fluxes of components $a$ (black; identified with the obscurer) and $c$ (red; identified with the warm absorber) normalized to their respective values at the first observational epoch.}
    \label{fig:ep1fluxnorms}
\end{figure}

Fig.~\ref{fig:ep1fluxnorms} shows the time variability for the fluxes of components $a$ (black circles; identified with the obscurer) and $c$ (red circles; identified with the warm absorber), with the fluxes normalized for each component to their flux at the first observational epoch. Whereas during most of the near-IR monitoring campaign the two components vary in opposite directions, as expected from the model of \citet{dehghanian19a}, there is a short period of about two months, where the two components seem to be synchronised. This phase indicates that density changes must also take place in the obscurer on relatively short time-scales, as already suggested by Fig.~\ref{fig:broadvdisk}. In particular, an increase in density would lead to an increase in absorption line equivalent width, which could counterbalance any decrease expected from an increasing X-ray flux alone, thus resulting in a net observed increase in line flux. Therefore, if the density is not constant, the He~{\sc i} equivalent width of the obscurer is not a good proxy for the incident X-ray radiation anymore. Therefore, in order to further test the Dehghanian et al. model we would need a near-IR campaign accompanied by X-ray monitoring. 

\subsection{The accretion disc wind in NGC~5548}

The complexity of the behaviour of various absorption components is not easy to accommodate in the global picture of NGC~5548 since some of the components are persistent, seen both in the IR and the UV, and responding to the change of the continuum (e.g. component $c$), while others, like components $b$ and $g$, do not display such behaviour, and in addition component $g$ implies inflow instead of outflow. All these narrow components, however, are persistent on timescales of at least a few months and must cover a significant fraction of the emission region in order to manifest as clear absorption features. Individual clouds from the BLR are too small to cover a significant part of the BLR as required by UV observations. They occasionally cause eclipses of the very compact X-ray source, but such events last of order of a few days or even less \citep[e.g.][]{Risaliti11,DeMarco20}. Clouds in the dusty, molecular torus can be larger since the density of the material there is expected to be lower, but the picture of the static torus itself is questionable, and winds are far more likely to provide the required obscuration when the source is seen at high viewing angles \citep{Elitzur06,Dorod12}.

\begin{figure*}
	\includegraphics[width=15cm]{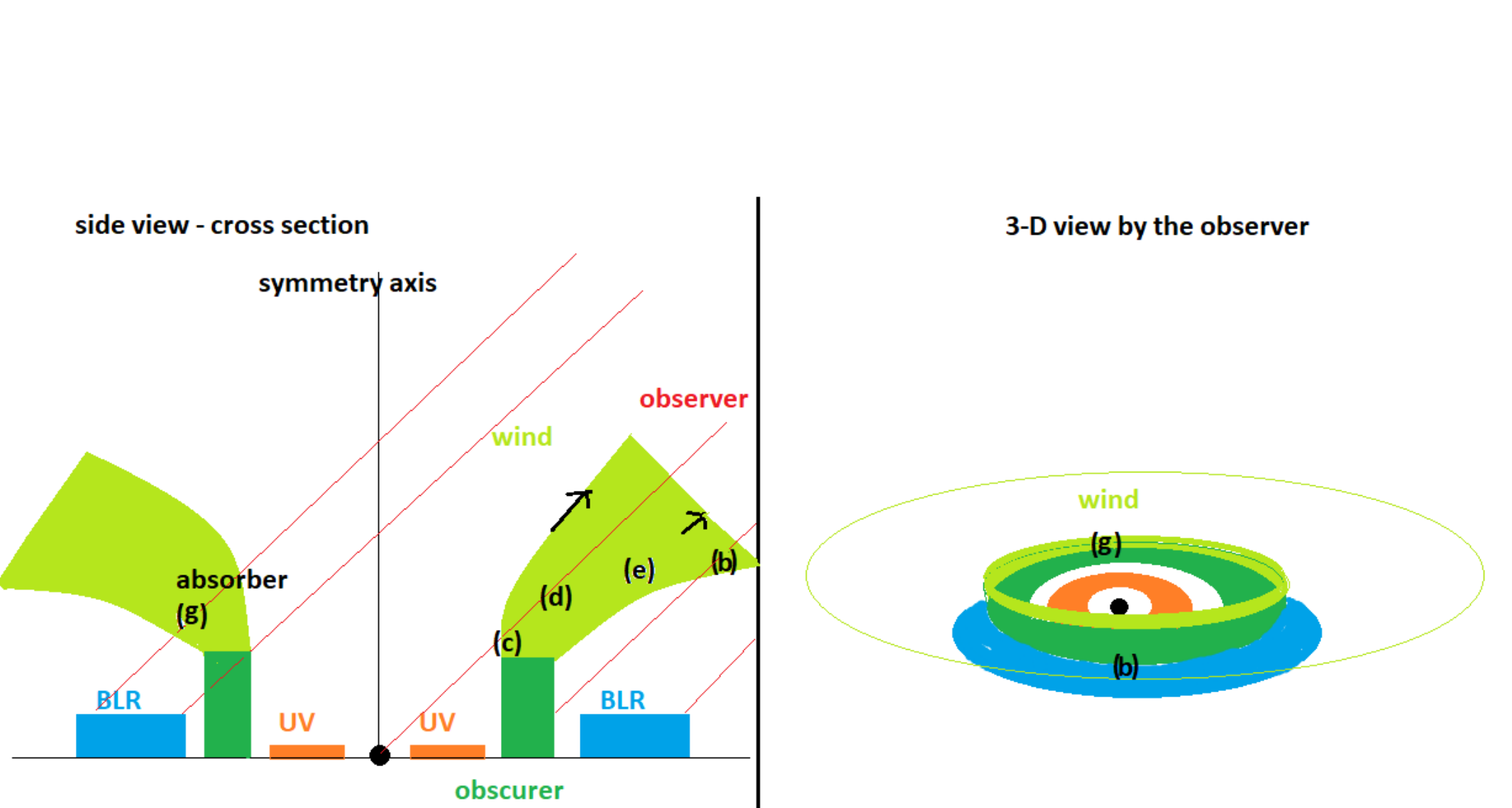}
    \caption{Our schematic picture of a generalised accretion disc wind in NGC~5548 assuming an inclination angle of 45$^\circ$ between the plane of the accretion disc (labelled ``UV'') and the observer's line of sight (labelled ``observer''). The accretion disc wind at the smallest radii in the plane (green) is dust-free and appears as the ``obscurer'' in its densest region (dark green), whereas the less dense regions further out are observed as the ``warm absorber''. The accretion disc wind at larger radii in the plane is dusty and achieves only a relatively small scale-height (Failed Radiatively Accelerated Dusty Outflow [FRADO]) observed as the broad-line region (labelled ``BLR'') at its illuminated face and as the dusty torus further out.}
    \label{fig:windschema}
\end{figure*}

\citet{dehghanian19b} discussed an accretion disc wind model for NGC~5548 that accounts for a number of observed properties, including the appearance of the obscurer, and in connection with this the broad and narrow ``line holidays''. We consider the geometry illustrated in their Fig.~1 as a framework within which to explain our new results. We build on this scenario as shown in Fig.~\ref{fig:windschema}, in order to provide a possible generalised schematic of the accretion disc wind. Our component $a$ is then related to the X-ray obscurer, identified as the base of the partially (and/or occasionally) translucent wind. Here the wind is expected to be the densest while also partially covering the observer's view of both the continuum region and the BLR.

In order to accommodate component $b$, which does not have a counterpart in the UV and is most likely a dust absorber, we must slightly modify the geometry of the wind. First, we postulate that the viewing angle with respect to the symmetry axis in this source is of order of 45$^\circ$, as recently determined by \citet{Horne20} on the basis of velocity-resolved reverberation mapping of the BLR during the 2014 HST campaign. The wind, initially launched perpendicular to the disc plane, bends under the radiation pressure and then propagates asymptotically roughly at an angle of 45$^\circ$ with respect to the symmetry axis \citep[see e.g.][]{Elvis00,Proga00,Matthews20}. If so, we observe the source under the conditions characteristic for broad-absorption line (BAL) quasars, partially looking through the flow. In NGC~5548, we do not see deep, extreme broad absorption features as in BAL quasars. This could be because the alignment is not perfect and/or the stream of the outflowing material is less dense. 

The observation of \citet{kriss19} of variability of the saturated UV broad absorption lines, which suggested covering factor changes in the obscurer, also fits well into this picture. In our interpretation, the WA components $c$, $d$, $e$ and $f$ are also located in this outflow, most likely in the less dense regions further out, and we observe the UV continuum emission and part of the BLR filtered by this material. The bending of the outflow after the ``obscurer'' region gives a natural explanation of why it can shield the X-ray continuum for the outer parts of the wind.

The opposite part of the outflowing wind in such a geometry has at its base the velocity vector at 90$^\circ$ with respect to the observer. Once the outflow partially bends towards the disc, we can see a small net velocity component in the opposite direction, imitating the inflow. Such a component (we consider this an attractive option for component $g$) would thus shield a considerable part of the BLR opposite with respect to the observer, but it would not shield the UV. The disappearance of the component at some times (e.g. it is not seen by \citet{kriss19}) must be related to the wind dynamics since the component does not respond in a simple way to the nuclear flux. However, this is possible since the wind material, moving with a speed of order of 3000~km~s$^{-1}$ travels the distance equal to the BLR radius over roughly a one-year timescale. The wind should be rather broad in the radial direction, bent, and with stratification of the velocity-wind streamlines launched at small radii will be faster than streamlines launched further out. In this way all absorption components can be accommodated in a single wind scenario.

The wind is likely located close to the onset point of the BLR itself since the asymptotic wind velocity should reflect the virial velocity at the launching radius. Therefore, its formation could be related to the formation of the BLR itself. An attractive scenario for the low-ionization part of the BLR is a dust-based model, either in a dynamical form of a Failed Radiatively Accelerated Dusty Outflow \cite[FRADO][]{Czerny11,Czerny17,Naddaf20} or in the form of a static puffed-up disc surface \citep{Baskin18}. The analytic considerations in \citet{Czerny15} showed that a dusty outflow is relatively the strongest at the inner radius of the BLR. But rather than dust launching the full model, we should consider also the radiation pressure from the lines which then leads to a sustained outflow, and not a failed wind. Possibly all these mechanisms are present in the general accretion disc model, dominating at different launch radii \citep[e.g.][]{Elvis12}.


\section{Conclusions}

In 2016/17, \citet{landt19} conducted a near-IR spectroscopic monitoring campaign on NGC~5548 and discovered He~{\sc i} $\lambda 10\,830$~absorption. We have decomposed this absorption into its components and have studied its time variability. Our main findings can be summarized as follows:\\

(i) We detect both narrow and broad absorption. We have decomposed the total He~{\sc i} absorption into seven components, with one of these identified with the obscurer that appeared in 2013. The appearance of He~{\sc i} narrow absorbers is most likely related to the appearance of the obscurer since this component is expected to reduce the ionization parameter of the warm absorber gas. \\

(ii) We detect significant time variability in both flux and equivalent width for the obscurer as well as the strongest narrow absorbers (in particular for the warm absorber component previously referred to as Component 1). We attribute the variability of the obscurer mainly to changes in the ionization parameter, however, density changes in its material may also be important (though to a lesser degree). The observed variability time-scales and behaviour support the recent interpretation of the obscurer as an accretion disc wind. \\

(iii) Since the He~{\sc i} absorption is unsaturated, we can use it to constrain the ionization parameter of the obscurer. Similar to \citet{kriss19}, we obtain ionization parameters of log~$\xi \sim 0.9$ and log~$\xi \sim 1.7$ for the obscurer components with hydrogen column densities derived from X-ray observations in 2013/14 of log~$N_{\rm H} \sim 22.1$~cm$^{-2}$ and log~$N_{\rm H} \sim 23.0$~cm$^{-2}$, respectively. \\

(iv) Since we directly observe the helium features from both the obscurer and the warm absorber, we can test the physical cycle of \citet{dehghanian19a} proposed to explain the ``line holiday'' in NGC~5548. As expected from their model, we find that during most of the near-IR monitoring campaign the two components vary in opposite senses. However, for a short duration the two components are synchronised, which suggests that changes in the density of the obscurer influence its He~{\sc i} flux in addition to the availability of soft-X-rays. \\

In the future, we plan to obtain high-spectral resolution near-IR spectroscopy of the He~{\sc i} absorbers in NGC~5548. Suitable instruments for such studies have recently become available, for example iShell on the IRTF and GIANO-B on the Telescopio Nazionale Galileo (TNG). In the longer term, we will pursue near-IR spectroscopy of AGN outflows quasi-simultaneous with spectroscopy at X-ray and UV frequencies. We believe that such an approach will yield new insights on important aspects of accretion disc winds. 


\section*{Acknowledgements}

C.W. and B.C. acknowledge funding from the National Science Center, Poland, through grant 2015/17/B/ST9/03436/ (OPUS 9). H.L., M.J.W. and D.K. acknowledge the Science and Technology Facilities Council (STFC) through grant ST/P000541/1. B.C. acknowledges further support by the National Science Centre, Poland, through grant 2017/26/A/ST9/00756 (Maestro 9).

\section*{Data availability}

The processed data underlying this work are available on request from the second author. The raw data are publicly available at the NASA IRTF Archive hosted by the NASA/IPAC Infrared Science Archive.




\bibliographystyle{mnras}
\bibliography{bib_cw} 

\bsp	
\label{lastpage}
\end{document}